\documentclass[times,final]{elsarticle}

\usepackage[margin=1in]{geometry}
\def\snm#1{#1}
\usepackage{framed,multirow}

\usepackage{amssymb}
\usepackage{latexsym}

\usepackage{physics}
\usepackage{bm}
\usepackage{subcaption}
\usepackage{bbold}
\usepackage{array}
\usepackage{mathtools}
\usepackage{pgfplots}
\DeclareUnicodeCharacter{2212}{−}
\usepgfplotslibrary{groupplots,dateplot}
\usetikzlibrary{patterns,shapes.arrows}
\pgfplotsset{compat=newest}
\usetikzlibrary{spy}
\usetikzlibrary{external}
\usepackage{url}
\usepackage{xcolor}
\definecolor{newcolor}{rgb}{.8,.349,.1}

\journal{Journal of Computational Physics}

\begin{document}

\begin{frontmatter}

\title{An immersed interface method for the 2D vorticity-velocity Navier-Stokes equations with multiple bodies}

\author[1]{James  \snm{Gabbard}}

\author[1]{Thomas \snm{Gillis}}
\author[2]{Philippe \snm{Chatelain}}

\author[1]{Wim M. \snm{van Rees}\corref{cor1}}
\cortext[cor1]{Corresponding author. E-mail address: wvanrees@mit.edu  
}

\address[1]{Department of Mechanical Engineering, Massachusetts Institute of Technology, 77 Massachusetts Avenue, Cambridge, MA 02139, USA} %
\address[2]{Institute of Mechanics, Materials and Civil Engineering, Université Catholique de Louvain, 1348 Louvain-la-Neuve, Belgium} %

\begin{abstract}

\noindent We present an immersed interface method for the vorticity-velocity form of the 2D Navier Stokes equations that directly addresses challenges posed by multiply connected domains, nonconvex obstacles, and the calculation of force distributions on immersed surfaces. The immersed interface method is re-interpreted as a polynomial extrapolation of flow quantities and boundary conditions into the obstacle, reducing its computational and implementation complexity. In the flow, the vorticity transport equation is discretized using a conservative finite difference scheme and explicit Runge-Kutta time integration. The velocity reconstruction problem is transformed to a scalar Poisson equation that is discretized with conservative finite differences, and solved using an FFT-accelerated iterative algorithm. The use of conservative differencing throughout leads to exact enforcement of a discrete Kelvin's theorem, which provides the key to simulations with multiply connected domains and outflow boundaries. The method achieves second order spatial accuracy and third order temporal accuracy, and is validated on a variety of 2D flows in internal and free-space domains.

\end{abstract}

\end{frontmatter}

\section{Introduction}
Immersed methods solve partial differential equations inside or outside of irregular domains, while using a regular structured grid (typically Cartesian).
The benefit of not having to adapt the underlying mesh to the domain boundaries provides simplicity and computational efficiency in handling complex domain geometries, arbitrary topologies (e.g.\ multiple immersed bodies), and dynamically moving domain boundaries. These characteristics are especially of interest when combined with the Navier-Stokes equations to solve flow problems such as biologically-inspired locomotion.
Broadly, there are two classes of immersed methods for incompressible Navier-Stokes simulations \citep{Mittal:2005}. Continuous forcing methods include traditional immersed boundary methods \citep{Peskin:1977, Mittal:2005, Taira2007},  and Brinkmann penalization \citep{Angot:1999, Coquerelle:2008, Gazzola:2011a, Gillis:2017}. These methods add a singular forcing term to the continuous Navier Stokes equations within solid regions, which approximately enforces a no-slip condition on solid boundaries. To maintain regularity after discretization, the forcing term is either smoothed on the object boundary and its value is resolved dynamically \citep{Gazzola:2011a}, or an iterative process is used to enforce the boundary condition \citep{Hejlesen:2015a,Gillis:2017}. This limits many such methods to first-order spatial and temporal accuracy. Discrete forcing methods, on the other hand, include sharp immersed boundary methods \citep{Mittal2008, Seo2011}, immersed interface methods \citep{Leveque1994, Li2006}, and other relatives such as Ghost Fluid \citep{Gibou2019}, Ghost Cell \citep{Tseng2003}, and cut cell finite volume methods \citep{Ingram2003}. These approaches use a modified discretization near solid objects that sharply resolves the location of immersed boundaries and enforces corresponding boundary conditions. Although these modifications are more challenging to derive and implement, they allow for increased spatial and temporal accuracy, as well as accurate resolution of local flow quantities such as traction forces on immersed solid boundaries.

Here we focus on the immersed interface method (IIM), a term that in itself covers a broad collection of discrete forcing methods. Early Navier-Stokes simulations used the IIM to discretize singular sources such as forcing terms representing interfaces with surface tension, or elastic membranes \citep{Li2001, Lee2003, Le2006}. With the introduction of the explicit jump immersed interface method (EJIIM) \citep{Wiegmann2000}, the IIM was extended from discretizing singular source terms to handle directly imposed boundary conditions such as Dirichlet or Neumann conditions. The EJIIM relies on the use of jump-corrected Taylor series within standard finite difference schemes, keeping the solution a linear combination of grid values while incorporating boundary conditions. Further, the method uses a dimensionally-split approach, simplifying its extension to 2D and 3D. Combined, the EJIIM and its newest iterations (including our work) overlap significantly with sharp immersed boundary methods, and have much in common with other methods (such as the Ghost Cell, Cut Cell, and Ghost Fluid Methods). 

Around the early 2000s, the IIM was combined for the first time with vorticity-based formulations of the 2D Navier-Stokes equations \citep{Calhoun2002, Li2003}, which used similar jump-corrected finite difference schemes as the EJIIM. These works are characterized by a temporal splitting approach to solve the Stokes problem with appropriate global vorticity boundary conditions, providing consistent second-order spatial accuracy and first-order temporal accuracy. In \citet{Linnick2005}, the authors employ the EJIIM to provide a 2D vorticity-velocity Navier-Stokes solver with compact difference schemes and a Thom-like, local vorticity boundary condition that enabled fourth-order spatial and temporal accuracy. This approach was recently extended using a more efficient multigrid solver in \citep{Hosseinverdi2018}, and implemented in 3D in \citep{Hosseinverdi2020}. Recently, the IIM has also been integrated within vortex particle-mesh methods, which rely on a combined Lagrangian-Eulerian approach to integrate the incompressible Navier-Stokes equations. In \citep{Marichal2014} a traditional Lighthill splitting approach was introduced to handle the vorticity boundary condition, leading to first-order temporal and second-order spatial accuracy. Subsequently, the same group employed a Thom-like finite difference boundary condition to achieve high-order accuracy in time \citep{Gillis2019}, and an extension to 3D \citep{GillisThesis}. These latter results used a Lattice Green's Functions FFT-accelerated Poisson solver, together with a Schur-complement boundary approach, solved using recycling GMRes \citep{Parks:2006,Gillis2018}.

The majority of the approaches above handle the external flow around a single, stationary, and typically convex object. A major challenge in extending towards multiple bodies is the need to enforce circulation conservation on each body independently. Here we build off the approaches in \citep{Linnick2005, Gillis2019} to develop a vorticity-based 2D finite-difference IIM that solves this issue, extending the scope of problems that can be simulated to bounded and unbounded fluid domains with multiple nonconvex immersed bodies and outflow boundary conditions. We use a conservative finite-difference discretization of the Navier-Stokes equations leading to second-order accuracy in space and third-order accuracy in time, with the elliptic system solved along the lines of \citep{Gillis2018}. Further, we consider for the first time in vorticity-velocity based IIMs  the calculation of time-dependent pressure and shear distributions on immersed surfaces. More broadly, we provide a novel interpretation of the EJIIM through the lens of ghost points reconstructed with a polynomial extrapolation, which greatly simplifies implementation details and exposes how the EJIIM is related to sharp-interface immersed-boundary and finite-volume methods.  

The rest of this work is structured as follows. Section \ref{section:iim} introduces the EJIIM and its relation to ghost point reconstruction, as well as its application to nonconvex bodies in 2D. Sections \ref{section:transport} and \ref{section:reconstruction} discuss IIM discretizations of the vorticity transport equation and the elliptic velocity reconstruction problem, respectively. In section \ref{section:navierstokes} these two discretizations are combined into a full Navier-Stokes discretization which enforces a discrete form of Kelvin's theorem. Section \ref{section:forces} introduces techniques for calculating forces and surface tractions acting on immersed bodies, which are applied to a variety of flows in section \ref{section:results} to illustrate the accuracy and effectiveness of the methods presented here. We conclude in section \ref{section:conclusion} with a summary of our contributions and a discussion of future directions for this work.
\section{The immersed interface method}\label{section:iim}
In this section we briefly review the explicit jump immersed interface method (EJIIM) in a 1D setting and discuss a specialization of the method which reduces complexity and enhances numerical stability. This specialization is then extended to 2D problems.

\subsection{Specializing the Explicit Jump IIM}\label{section:simpleEJIIM}

\begin{figure}
    \centering
    \begin{subfigure}{0.49\linewidth}
        \centering
        \includegraphics[width=\textwidth]{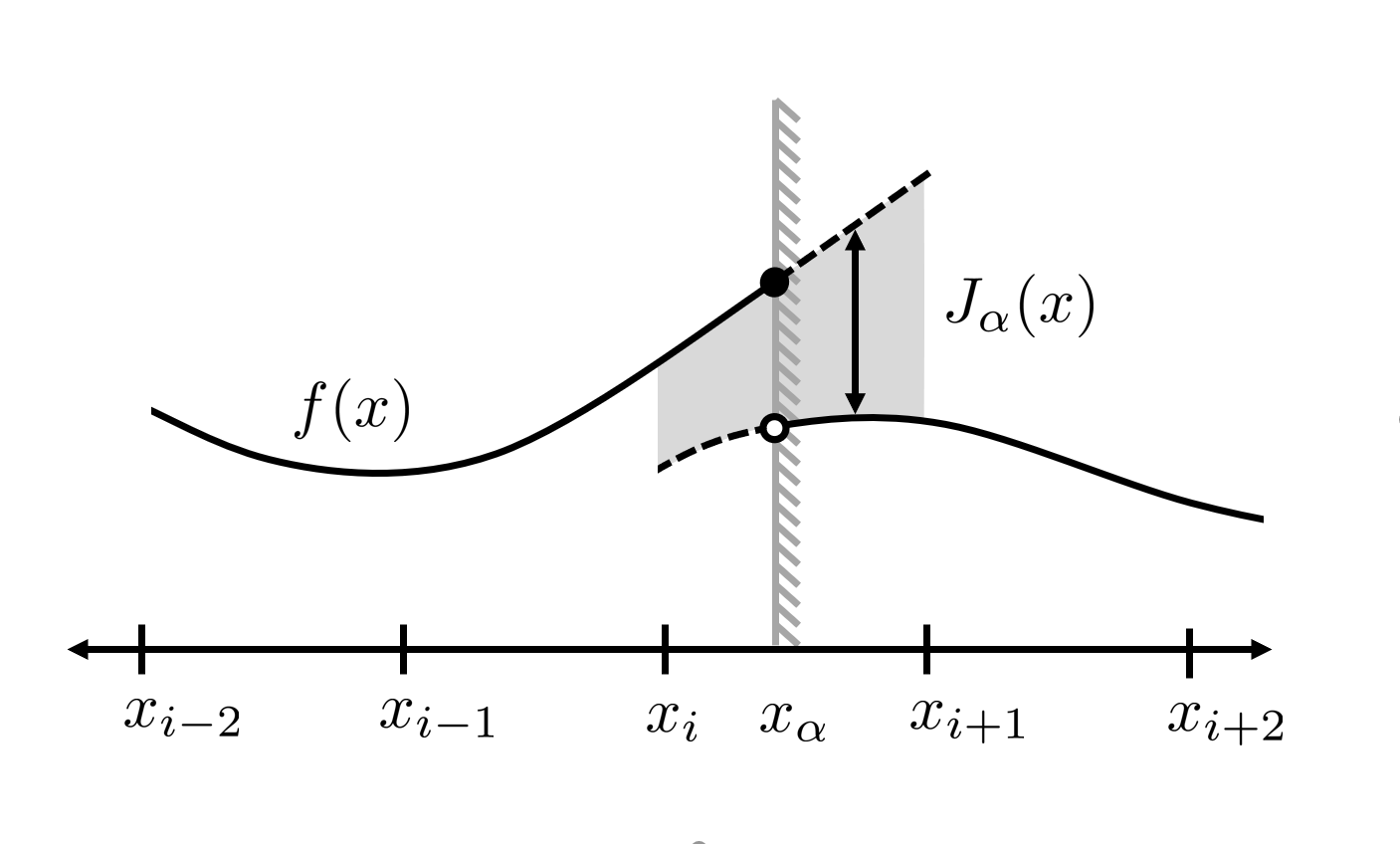}
        \caption{}
        \label{fig:EJIIM}
    \end{subfigure}
    \begin{subfigure}{0.49\linewidth}
        \centering
        \includegraphics[width=\textwidth]{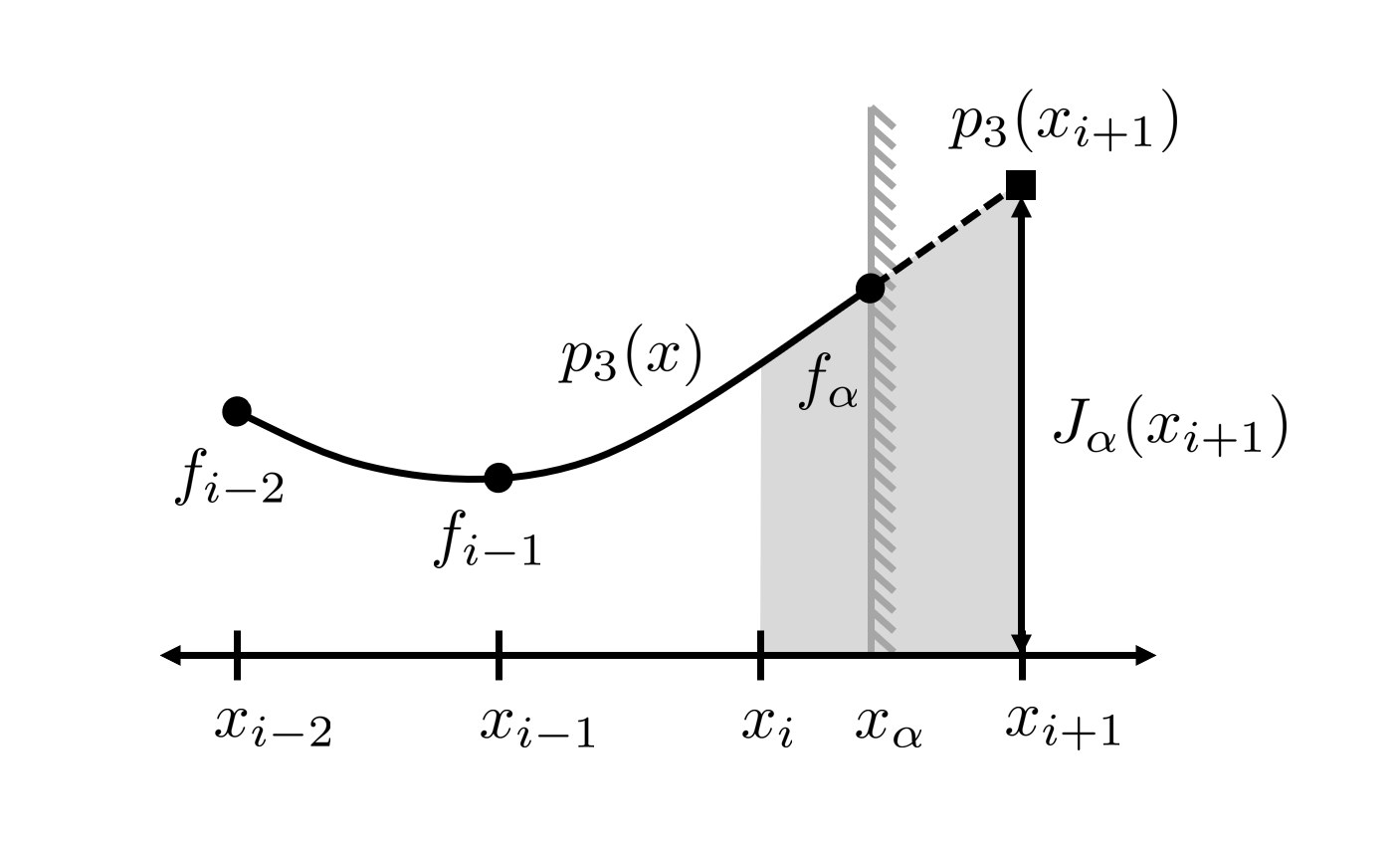}
        \caption{}
        \label{fig:simplerEJIIM}
    \end{subfigure}
    \caption{Two use cases for the Explicit Jump IIM. (a) For a physical interface, the solution $f(x)$ has nontrivial dynamics on both sides of a discontinuity. The jump in $f(x)$ and its derivatives are known a priori from physical principles. (b) For a domain boundary, the solution is trivial on one side of the discontinuity, and a Dirichlet boundary condition is prescribed. However, there is no information on the jump in the derivatives.}
\end{figure}

The explicit jump immersed interface method (EJIIM), introduced by \citet{Wiegmann2000}, is a method of adapting regular finite difference schemes for equations with discontinuous solutions. At its center is the construction of modified Taylor series expansions which correctly approximate functions with jump discontinuities. To illustrate, consider a function $f(x)$ that is smooth except at a point $x_\alpha$, where there is a jump singularity in $f(x)$ and its derivatives $f^{(k)}(x)$. Let $(f^{(k)})^-$ and $(f^{(k)})^+$ denote the value of $f^{(k)}$ on the left and right sides of the discontinuity, respectively, and let $[f^{(k)}]_\alpha = (f^{(k)})^+ - (f^{(k)})^-$ denote the magnitude of the jump in $f^{(k)}$ at $x_\alpha$. Finally, consider a regular grid of points $x_i = ih$, with the point $x_\alpha$ contained in the interval $[x_i, x_{i+1}]$ (Figure \ref{fig:EJIIM}). Given the values of $f^{(k)}(x_i)$ and $[f^{(k)}]_\alpha$, the function $f$ can be extrapolated from $x_i$ to $x_{i+1}$ using the modified Taylor series
\begin{equation}\label{jumpcoeff}
        f(x_{i+1}) = \sum_{k=0}^{n} \frac{h^k}{k!}f^{(k)}(x_i)  + J_\alpha(x_{i+1}) + O(h^{n+1}), \quad\text{with}\quad
        J_\alpha(x_{i+1}) =\sum_{k=0}^{n} \frac{(x_{i+1} - x_\alpha)^k}{k!}[f^{(k)}]_\alpha.
\end{equation}
The first half of (\ref{jumpcoeff}) is a standard Taylor expansion of $f$ about $x_i$; the second is a jump correction that must be added to any expansion that crosses the discontinuity. In the EJIIM, these generalized Taylor series are used to construct jump-corrected finite difference stencils which retain their high-order accuracy across the jump discontinuity at $x_\alpha$. This method is well suited to physical interfaces where the jumps $[f^{(k)}]_\alpha$ are determined by the geometry of the interface or a known discontinuity in a prescribed source field.

Since its original publication, the EJIIM has been repeatedly re-purposed to discretize problems with smooth solutions that are posed on irregular domains. To tackle such problems, a common approach is to prescribe the solution outside of the problem domain (typically to zero value), and then treat the irregular domain boundary as a jump discontinuity. In this case the jump discontinuity is no longer physically constrained, and the value of the jump in each derivative $f^{(k)}(x_\alpha)$ must be calculated directly from the function $f(x)$ by evaluating a one-sided finite difference stencil \citep{Linnick2005, Marichal2014, Gillis2019}. To illustrate this procedure, consider the same function $f$ discussed above, now with $f(x) = 0$ for $x > x_\alpha$ to model a domain boundary (Figure \ref{fig:simplerEJIIM}). Given the value of $f(x)$ on the regular grid and a Dirichlet boundary condition $f(x_\alpha)$, the interface derivatives $(f^{(k)})^-$ are approximated by the $n+1$ point one-sided finite differences
\begin{equation}\label{eq:wallstencil}
    (f^{(k)})^-_{FD} = S^k_\alpha f(x_\alpha) + \sum_{j=1}^{n} S^k_j f(x_{i-j})  = (f^{(k)})^- + \order{h^{n+1-k}}.
\end{equation}
Here the point $x_i$ has been excluded from the stencil to avoid ill conditioning when $\abs{x_\alpha - x_i}$ is small. The optimal order of accuracy in \eqref{eq:wallstencil} is achieved by taking $S_j^{(k)} = \ell_j^{(k)}(x_\alpha)$, where the $\ell_j(x)$ are the degree $n$ Lagrange polynomials satisfying $\ell_i(x_j) = \delta_{ij}$ for $x_j \in \{x_\alpha, \; x_{i-1}, \; x_{i-2}, \;..., \; x_{i-n}\}$. Using the approximation of $(f^{(k)})^-$ given in \eqref{eq:wallstencil} and taking $(f^{(k)})^+ = 0$ leads to the approximate jump correction
\begin{equation}\label{eq:onesided}
     J_\alpha(x_{i+1}) = - \sum_{k=0}^n \frac{(x_{i+1} - x_\alpha)^k}{k!} (f^{(k)})^-_{FD} + \mathcal{O}\qty(h^{n+1}),
\end{equation}
which is the expression used in \citep{Marichal2014, Marichal2016, Gillis2018, Gillis2019}.

In this work we simplify the evaluation of $J_\alpha(x_{i+1})$ by defining the degree $n$ interpolating polynomial $p_n(x) = \sum_j f(x_j) \ell_j(x)$ and noting that $(f^{(k)})^-_{FD} = p_n^{(k)}(x_\alpha)$. Making this substitution in \eqref{eq:onesided} reveals that $J_\alpha(x_{i+1})$ is an $(n+1)$ term Taylor expansion of $-p_n(x)$ about the point $x_\alpha$, which is equivalent to the evaluation
\begin{equation}
    J_\alpha(x_{i+1}) = -p_n(x_{i+1}).
\end{equation}
Thus the evaluation of $n$ one-sided finite differences can be replaced with a single evaluation of the interpolating polynomial via Neville's algorithm \citep{Press2007}. This implies that at domain boundaries the jump-corrected finite differences of the EJIIM are equivalent to a polynomial extrapolation of $f(x)$ followed by the evaluation of a standard finite difference stencil, which closely links our specialization of the EJIIM to other extrapolation-based procedures such as sharp immersed boundary methods \citep{Mittal2008, Seo2011} and the Ghost Fluid method \citep{Gibou2019}.

This result generalizes to a variety of situations beyond the one described above. If $f(x)$ is nonzero for $x > x_\alpha$ due to an uncoupled physical process occurring on the other side of the interface, then the contributions to $J_\alpha(x_{i+1})$ from $(f^{(k)})^+$ can be approximated by an evaluation of $f(x)$:
\begin{equation}
    J_\alpha(x_{i+1}) = f(x_{i+1}) - p_n(x_{i+1}) + \order{h^{n+1}}.
\end{equation}
If a Neumann condition is present at $x_\alpha$ instead of a Dirichlet condition, then $p_n(x)$ is replaced by the unique interpolating polynomial satisfying $p^{(1)}(x_\alpha) = f^{(1)}(x_\alpha)$ and interpolating $f(x)$ at the $n$ neighboring points $\{x_{i-1}, \;..., \; x_{i-n}\}$. If no boundary condition is available, then the boundary point $x_\alpha$ is replaced by $x_i$ in the interpolation stencil. All of these specializations of the EJIIM reduce complexity and computational cost. They also improve numerical stability, by eliminating the need to invert a Vandermonde matrix when calculating one-sided stencil coefficients. In the remainder of this work, we therefore use this reduction of the EJIIM through polynomial extrapolation.

\subsection{Extending the IIM to 2D and nonconvex geometries}\label{section:2diim}
\begin{figure}
    \centering
    \includegraphics[width=0.7\textwidth]{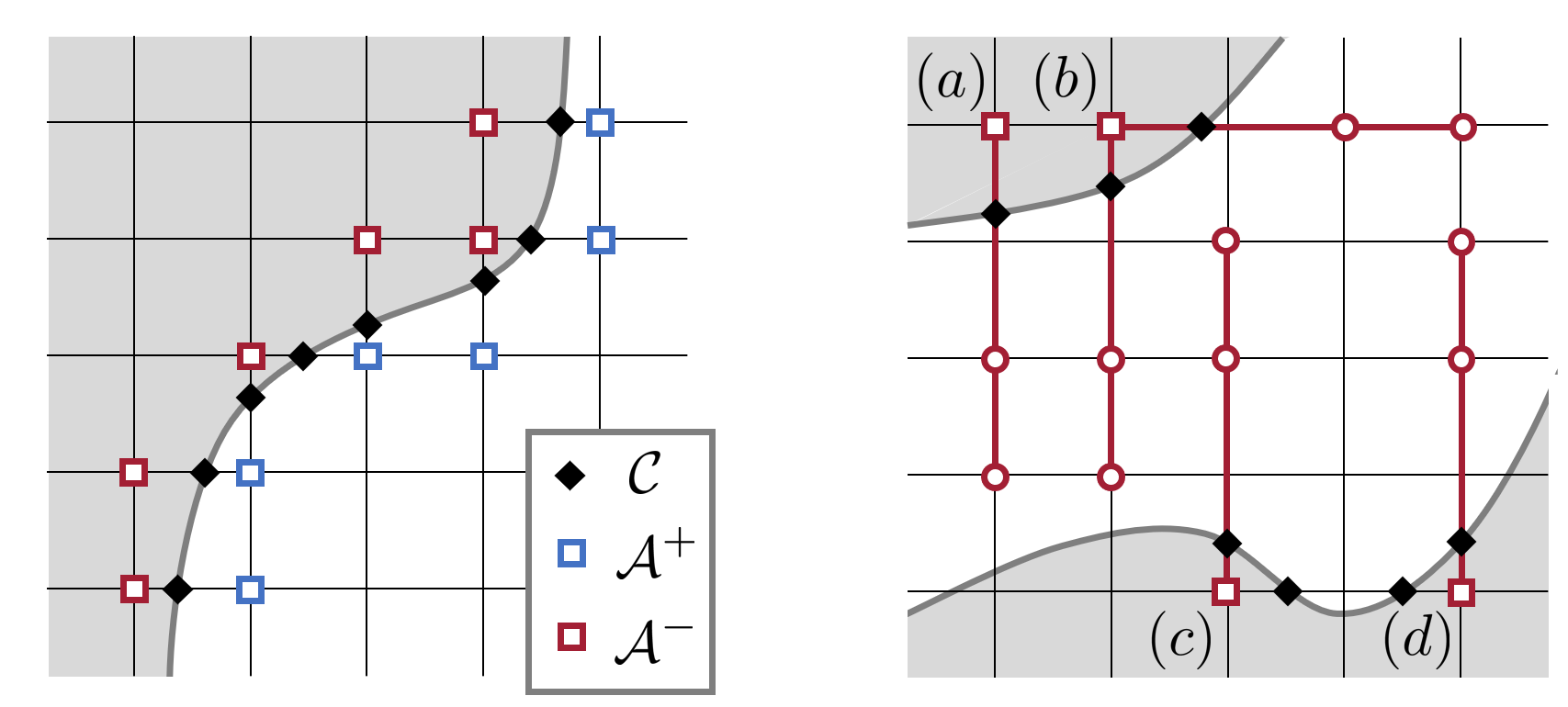}
    \caption{(left) The control points $\mathcal{C}$ and affected points $\mathcal{A}$ for a typical IIM discretization. The grey region represents points lying outside the problem domain. (right) Stencils used to fill the affected points $\mathcal{A}^-$ with a third order extrapolation. Point (a) has only one neighbor in the problem domain, and consequently receives an extrapolation along only one coordinate direction. Point (b) has two neighbors, so its value is the average of two separate one-dimensional extrapolations. Points (c) and (d) have two neighbors in the problem domain, but there are not enough points in the domain to allow for a third-order horizontal extrapolation stencil; consequently, each is filled from a vertical extrapolation only.}
    \label{fig:2dnotation}
\end{figure}

The sharp-interface method inspired by the EJIIM, as described above, extends readily to multidimensional problems, where the polynomial extrapolation perspective further helps address the challenges associated with concave geometry. To discuss multidimensional immersed interface methods, we define some convenient terminology and notation. Let $\mathcal{G}$ be the set of all nodes of a uniform Cartesian grid, and let $\Omega$ be an irregular two-dimensional domain immersed in this grid. Following the lines of \citet{Gillis2018}, we define two additional sets of points:
\begin{itemize}
    \item The set of \textbf{Control points} (denoted by $\mathcal{C}$) contains all intersections between the Cartesian grid and the immersed boundary $\partial \Omega$. 
    \item The set of \textbf{Affected points} (denoted by $\mathcal{A}$) are regular grid points that are adjacent to a control point. $\mathcal{A}$ can be subdivided into two disjoint sets $\mathcal{A}^+$ and $\mathcal{A}^-$, which represent points that are inside and outside $\Omega$, respectively.  
\end{itemize}
The sets $\mathcal{C}$, $\mathcal{A}^-$, and $\mathcal{A}^+$ are labeled in Figure \ref{fig:2dnotation} for a typical domain. 
\ref{section:geometryprocessing} provides a method for efficiently calculating the set of control points $\mathcal{C}$ for smooth geometries represented by a level set. With the control points found, our 2D immersed interface method precedes each finite-difference evaluation by a polynomial extrapolation which gives nonzero values to the inner affected points $\mathcal{A}^-$. For inner affected points with only one neighboring control point, this value is determined using a one-dimensional polynomial extrapolation along the associated grid line (as discussed in section \ref{section:simpleEJIIM}). This is equivalent to the dimension-splitting techniques implemented in \citep{Gillis2018,Gillis2019}. For inner affected points with multiple neighboring control points, we follow \citet{Marichal2016} by averaging the results of the 1D extrapolations along associated grid lines.
Both of these cases are illustrated in Figure \ref{fig:2dnotation}. The averaging procedure preserves the order of accuracy of the extension, and provides a minimal way to reconcile 1D extrapolations taken from different directions.

For nonconvex geometries, some control points may not have enough immediate neighbors to form an interpolating polynomial of the correct degree. This challenge does not disappear as the grid is refined: any amount of concavity, no matter how slight, can lead to control points with as few as one immediate neighbor (Figure \ref{fig:2dnotation}). 
One way to avoid this is to rely on extrapolations taken from multiple coordinate directions, as discussed above. This strategy is suggested and successfully implemented by \citet{Hosseinverdi2018} for an immersed interface method based on compact finite differences. In the method presented here, control points with too few neighbors are simply ignored, and the associated points in $\mathcal{A}^-$ are filled using an extrapolation along a different coordinate direction. For smooth geometries, this extrapolation method allows every point in $\mathcal{A}^-$ to be filled given sufficient resolution. Nonsmooth geometries present additional challenges, including cusps and acute interior corners, which are left for future work.

\section{Vorticity Transport}\label{section:transport}
We now proceed with combining our IIM with the discretization of transport equations in 1D and 2D. Specifically, we focus on the vorticity evolution equation governing the 2D incompressible Navier-Stokes equation, written here in conservative form:
\begin{equation}\label{eq:diffconservation}
    \pdv{\omega}{t} + \nabla \cdot (\vb{u}\omega - \nu \nabla \omega) = 0.
\end{equation}
Integrating this differential conservation law over a 2D region $R$ leads to the integral form
\begin{equation}\label{eq:intconservation}
    \dv{}{t} \int_R \omega \dd{A} + \oint_{\partial R} (\vb{u}\omega - \nu \nabla \omega) \cdot \vu{n} \dd{s} = 0.
\end{equation}
In this section, we discretize \eqref{eq:diffconservation} with a conservative finite difference scheme, using numerical fluxes as described in \citet{Shu1998}. This leads to a discrete form of the integral conservation law \eqref{eq:intconservation}, which is essential to the discretization of Kelvin's theorem presented in section \ref{section:kelvin}. We also develop an immersed interface boundary treatment that respects the mixed hyperbolic-parabolic character of \eqref{eq:diffconservation}, and show that it does not degrade the stability or accuracy of the free-space scheme. 

\subsection{Free space discretization}
For illustration, we begin with the one-dimensional advection-diffusion equation 
\begin{equation}\label{eq:1dtransport}
    \pdv{\omega}{t} + \pdv{}{x}\qty(u\omega - \nu \pdv{\omega}{x}) = 0,
\end{equation}
with a spatially varying velocity $u$ and constant diffusivity $\nu$. Let $f = u\omega$ be the advective flux, and $q = -\nu \omega_x$ be the diffusive flux, so that $\omega_t + (f + q)_x = 0$. To discretize this equation, consider a one-dimensional grid of points $x_i$ with regular spacing $h$, along with a discrete vorticity field $\omega_i$ and velocity field $u_i$. If numerical fluxes $f_{i + \frac 12}$ and $q_{i + \frac 12}$ are defined at the half grid points $x_{i + \frac 12} = \frac{1}{2}(x_i + x_{i+1})$, then \eqref{eq:1dtransport} can be approximated with centered differences:
\begin{equation}\label{eq:shuconserve}
    \dv{\omega_i}{t} + \frac{f_{i + \frac 12} - f_{i - \frac 12}}{h} + \frac{q_{i + \frac 12} - q_{i - \frac 12}}{h} = 0.
\end{equation}
 The spatial accuracy of the conservative discretization depends on the interpolation procedure used to construct $f_{i + \frac 12}$ and $q_{i + \frac 12}$. Here we choose a third-order upwind advective flux using the stencils \citep{Shu1998}
\begin{equation}\label{upwind}
    f_{i+\frac 12} = \max(u_{i+\frac 12},0) \cdot \qty[-\frac{1}{6} f_{i-1} + \frac{5}{6} f_i + \frac{1}{3} f_{i+1}] + \min(u_{i+\frac 12},0)\cdot \qty[\frac{1}{3}f_i + \frac{5}{6} f_{i+1} -\frac{1}{6} f_{i+2}],
\end{equation}
where $f_i = u_i \omega_i$, and the local upwind direction is determined by  $u_{i+\frac 12} =(u_i + u_{i+1})/2$. 
The diffusive flux $q_{i+\frac 12}$ is discretized with the centered difference 
\begin{equation} 
    q_{i+\frac 12} = -\nu\frac{\omega_{i+1} - \omega_i}{h},
\end{equation}
which leads to overall second order accuracy for the diffusion term.

The 1D conservative transport discretization developed above is readily extended to two dimensions. For a 2D transport equation with velocity field $\vb{u} = (u_x, u_y)$, define the $x$-direction fluxes $f_x = u_x\omega$ and $q_x = -\nu\dv{\omega}{x}$ at the $x$-direction flux points $\vb{x}_{i+1/2, j} = (x_{i+1/2}, y_j)$, and the $y$-direction fluxes $f_y = u_y \omega$ and $q_y = -\nu\dv{\omega}{y}$ at the $y$-direction flux points $\vb{x}_{i, j+1/2} = (x_i, y_{j+1/2})$. Each of these fluxes is calculated by applying the one-dimensional schemes along the corresponding grid line, and the full transport equation is discretized using the differencing scheme 
\begin{equation}\label{eq:2dshuconserve}
    \dv{\omega_{i,j}}{t} + \frac{f_{i + 1/2,j} - f_{i - 1/2,j}}{h} + \frac{f_{i,j+1/2} - f_{i,j - 1/2}}{h} + \frac{q_{i + 1/2,j} - q_{i - 1/2,j}}{h} + \frac{q_{i,j + 1/2} - q_{i,j - 1/2}}{h}= 0.
\end{equation}
This 2D transport scheme obeys a discrete form of the integral conservation law \eqref{eq:intconservation}. Because a similar discrete conservation property appears in the velocity reconstruction problem (section \ref{section:iimpoisson}) and in the enforcement of Kelvin's theorem (section \ref{section:kelvin}), we define notation for it here. Let $R$ be a 2D rectangular region with boundaries passing through the flux points, enclosing the set of grid points $\{\vb{x}_{ij}: \ell_x \le i \le r_x, \; \ell_y \le j \le r_y\}$. The total vorticity in $R$ can be approximated with the second-order quadrature
\begin{equation}
\label{eq:sum_R_def}
 \int_R\omega \dd{A} \approx h^2 \sum_{i = \ell_x}^{r_x}  \sum_{j = \ell_y}^{r_y} \omega_{i,j} \equiv h^2 \sum_R \omega,
\end{equation}
where we define the discrete operator $\sum_R (\cdot)$ as the sum over grid values within the region $R$. Similarly, a numerical flux $\vb{f}$ can be integrated over the boundary of $R$ with the second-order discrete contour integral
\begin{equation}
\label{eq:sum_dR_def}
\oint_{\partial R} \vb{f} \cdot \vu{n} \dd{s} \approx h \sum_{j = \ell_y}^{r_y} f_{r_x+1/2, j} + h \sum_{i = \ell_x}^{r_x} f_{i, r_y+1/2}  - h \sum_{j = \ell_y}^{r_y} f_{\ell_x-1/2, j} - h \sum_{i = \ell_x}^{r_x} f_{i, \ell_y-1/2} \equiv h \sum_{\partial R} \vb{f} \cdot \vu{n}, 
\end{equation}
where the four single summations represent the flux across the left, top, right, and bottom faces respectively, and $\sum_{\partial R}$ is shorthand for the sum over all these faces. Using this notation, the telescoping sum property of \eqref{eq:2dshuconserve} leads to the discrete conservation law satisfied by our numerical scheme 
\begin{equation}\label{2dintconserve}
\dv{}{t} \qty( h^2\sum_R  \omega) + h\sum_{\partial R} (\vb{f + q}) \cdot \vu{n} = 0,
\end{equation}
which approximates the continuous integral conservation law to second order. This relation could be easily extended to more complex grid-aligned regions if needed, by noting that any such region can be written as a union of grid-aligned rectangles.

The stability of this free-space transport discretization is discussed in \ref{section:transportstability}; the scheme is conditionally stable when integrated with a second or third order Runge-Kutta scheme.

\subsection{Immersed interface boundary treatment}
For finite domains, the advection-diffusion equation \eqref{eq:diffconservation} requires a single boundary condition for $\omega$ on each boundary. Here we consider the case where Dirichlet boundary conditions $\omega_b$ and $\vb{u}_b$ are known in advance, which will be most relevant for a discretization of the full Navier-Stokes equations. Because of the distinct parabolic and hyperbolic nature of the diffusion and advection terms, the Dirichlet boundary conditions are handled differently when discretizing each.

Let the grid points $x_1$ to $x_N$ form a finite computational domain, with immersed boundaries at $x_\ell \in [x_0, x_1]$ and $x_r \in [x_N, x_{N+1}]$. To calculate the diffusive flux on a domain with immersed boundaries, the scalar field $\omega$ is extrapolated to $x_0$ and $x_{N+1}$ using fourth-order polynomial extrapolations that depends on the Dirichlet boundary conditions $\omega_b$, as described in section \ref{section:simpleEJIIM}. Once this is done, the calculation of diffusive fluxes proceeds as usual for all flux points between $x_{1/2}$ and $x_{N+1/2}$. The use of a fourth-order extension leads to second-order accuracy for the diffusive term right up to the immersed boundary. This process is extended to 2D domains in a completely analogous way: the vorticity field is extrapolated to the inner affected points $\mathcal{A}^-$ at fourth order, using the Dirichlet boundary condition, and afterwards the diffusive fluxes are calculated normally. 

The calculation of the advection term requires a different procedure that addresses its hyperbolic nature. For a purely hyperbolic equation, boundary conditions are only necessary on regions of the boundary which act as an inflow. The use of a boundary condition on outflow boundaries leads to ill-posed continuous problems and instability in numerical discretizations. Consistent with this, we extrapolate the vorticity field past each domain boundary at third order, using the Dirichlet condition at inflow boundaries and ignoring the Dirichlet condition at outflow boundaries. 
The width of the third-order upwind advection stencil, which extends two points beyond each inflow boundary, presents an additional challenge. To avoid extra extrapolation, we force the choice of a downwind stencil at inflow boundaries, which extends only one point beyond the boundary. This strategy maintains the overall second-order accuracy of the transport discretization and does not affect the observed stability of the method. 

\begin{figure}
    \centering
    \includegraphics[width=0.7\textwidth]{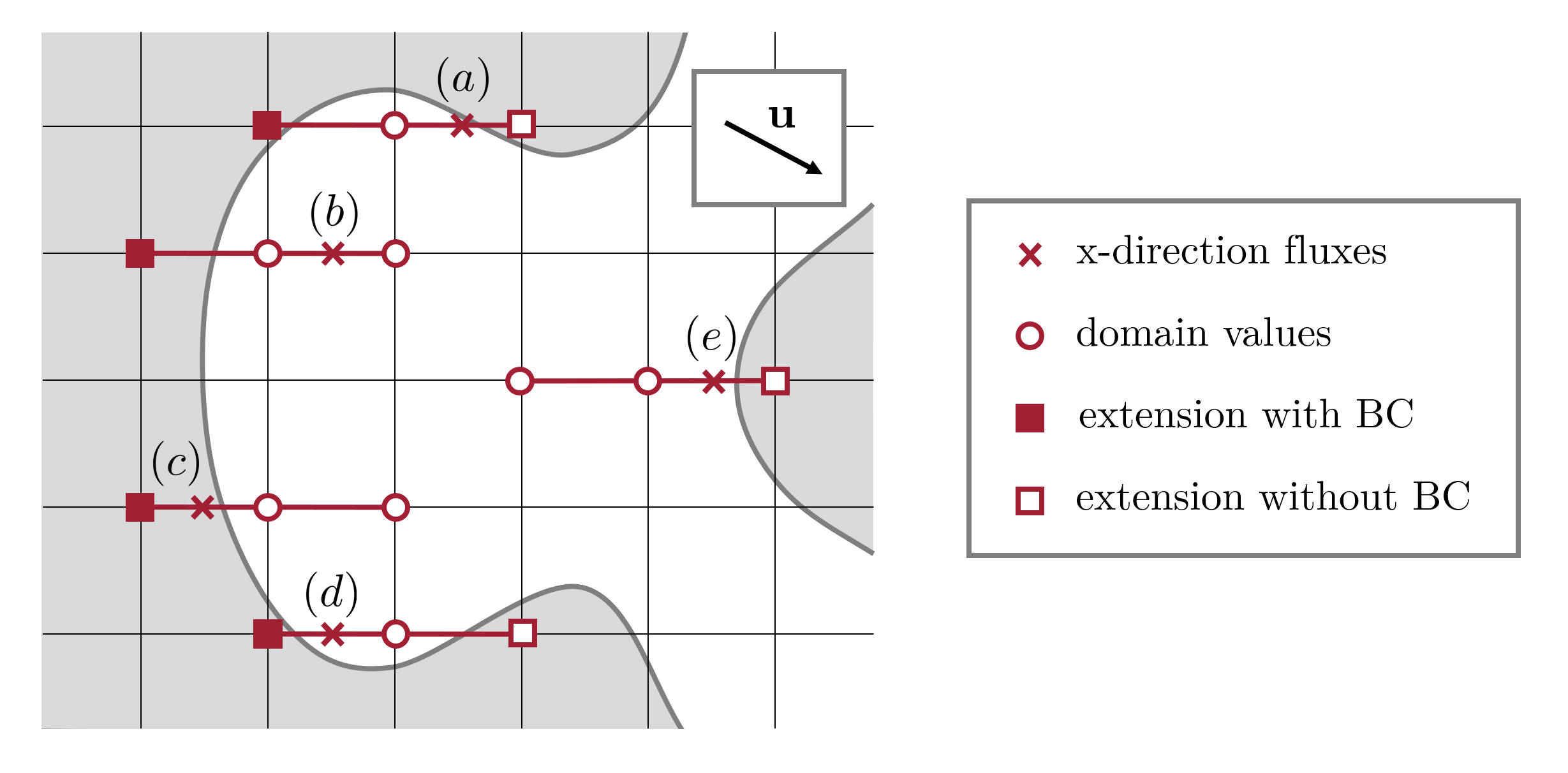}
    \caption{Boundary treatment for the third-order upwind advective flux in 2D. Here the upwind direction for x-direction fluxes is to the left, so that boundary fluxes (c) and (d) are treated as inflows while (a) and (e) are treated as outflows. Fluxes within the domain (b) and at the outflow boundaries (a, e) are calculated with the standard upwind stencil. Fluxes at the inflow boundaries (c, d) use a downwind stencil. In all cases, whenever a stencil crosses the boundary, an extension with boundary condition fills the upwind end (a, b, c, d), while an extension without boundary condition fills the downwind end (a, d, e).}
    \label{fig:2dflux}
\end{figure}

The 1D boundary treatment described above is readily extended to 2D domains. The velocity field is extended to $\mathcal{A}^-$ at third order, using the boundary velocity $\vb{u}_b$. The vorticity field is extended to $\mathcal{A}^-$ twice at third order, once with the Dirichlet condition $\omega_b$ and once without, so that both sets of values are available for flux calculations. Afterwards, the local upwind direction at each flux point is determined from the extended velocity field. In the fluid domain, each advective flux is calculated with a one-dimensional third order upwind stencil, using the extended vorticity field (with boundary condition $\omega_b$) where necessary (Figure \ref{fig:2dflux}b). On the boundary, each one-dimensional flux that acts as a local inflow is calculated with a downwind stencil (Figure \ref{fig:2dflux}c and \ref{fig:2dflux}d), while each flux that acts as a local outflow is calculated with an upwind stencil (Figure \ref{fig:2dflux}a and \ref{fig:2dflux}e). Regardless of their orientation, all boundary fluxes use the vorticity field extended with boundary condition on the upwind end of their stencil, and the vorticity field extended without boundary condition on the downwind end of their stencil; see Figure \ref{fig:2dflux} for illustration. 

\subsection{Numerical results}\label{section:transportconvergence}
\begin{figure}[ht]
    \centering
    \begin{minipage}{0.4\textwidth}
    \includegraphics[width=\textwidth]{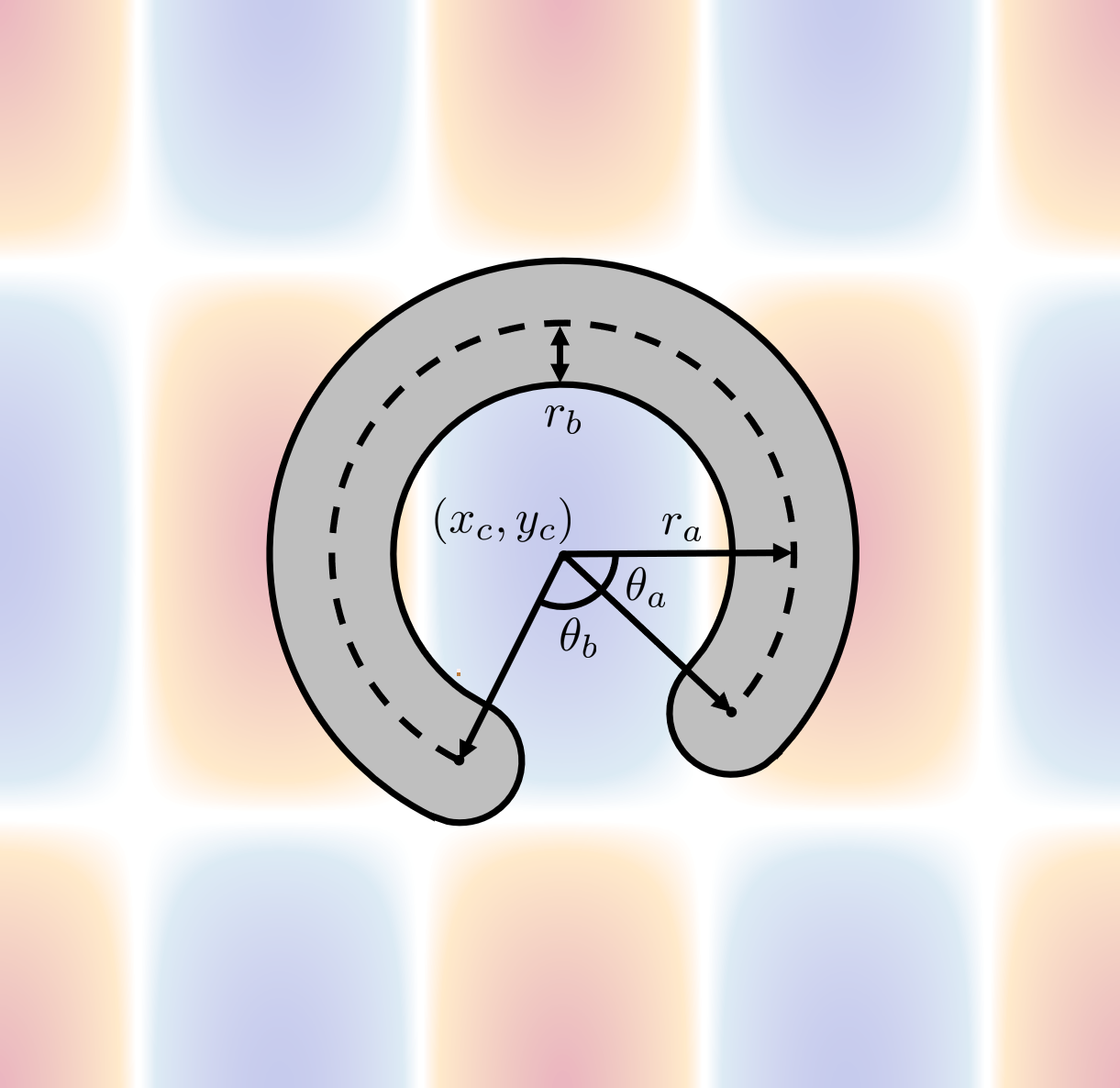}
    \end{minipage}
    \hspace{0.04\textwidth}
    \begin{minipage}{0.45\textwidth}
            \centering
            \begin{tabular}{ | p{0.25\linewidth} | p{0.7\linewidth} | } 
            \hline
            Element & Parameters \\ 
            \hline\hline
            Grid & $\vb{x}_{ij} = (i/N, \; j/N)$ for $0 \le i,j \le N - 1$. \\
            \hline
            Time steps & $T = 1.0, \delta t = 0.9 \Delta t_{max}$ \\
            \hline 
            Solution & $\vb{u} = (1.0,2.0)$, $\nu = (0.002, 0.064)$, $\vb{k} = (4\pi, 2\pi)$ \\
            \hline 
            Solid body & $\vb{x}_c = (0.507, 0.531)$, $r_a = 0.201$, $r_b = 0.054$, $\theta_a = 0.5$, $\theta_b = 2.4 $ \\
            \hline
            \end{tabular}
    \end{minipage}
    \caption{Geometry and parameters used in the transport convergence study.}
    \label{fig:transportsetup}
\end{figure}

To demonstrate the stability and convergence of the transport scheme proposed above, we consider a 2D test case with a known solution. A non-convex solid body is superimposed on the spatially periodic vorticity field
\begin{equation}
    \omega_{ex}(\vb{x}, t) = \cos [k_x(x - u_x t)] \cos [k_y(y - u_y t)] e^{-\nu k^2 t},
\end{equation}
which is an exact solution to the vorticity transport equation for constant velocity $\vb{u} = (u_x, u_y)$, viscosity $\nu$, and wavenumber $\vb{k} = (k_x, k_y)$. The boundary conditions $\omega_b$ and $\vb{u}_b$ on the solid body are set to match this solution, and a periodic boundary condition is prescribed on the edge of the computational domain. The flow is discretized on a square grid with $N$ points along each side, and integrated from $t = 0$ to $t = T$ using a third order Runge-Kutta method. The time step $\Delta t$ is chosen to be $0.9 \Delta t_{max}$, where $\Delta t_{max}$ is the maximum stable time step for the transport scheme (as determined by the procedure in \ref{section:transportstability}). Figure \ref{fig:transportsetup} defines the geometry of the solid body and lists the exact discretization parameters used in this test case.

\begin{figure}[h]
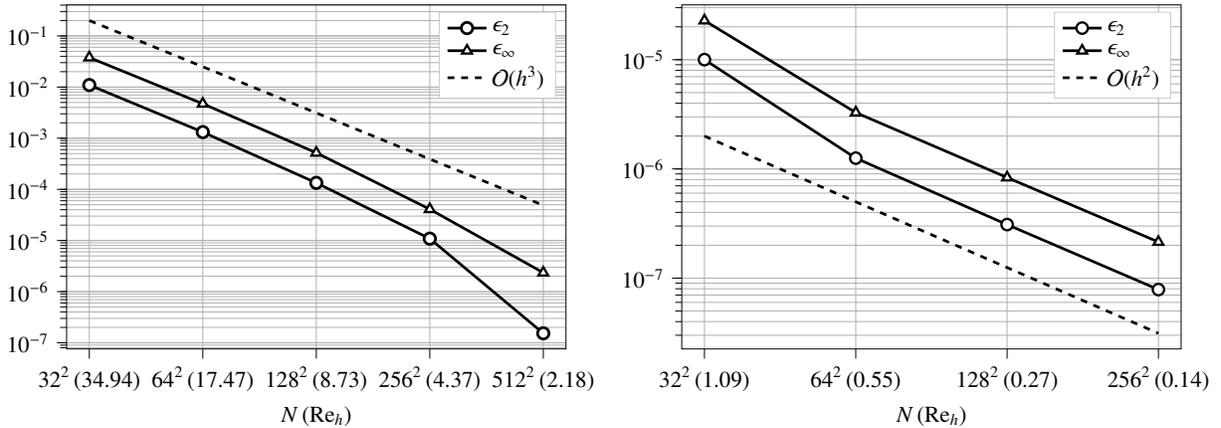

    \centering
    \begin{subfigure}{0.49\textwidth}
        \centering
        \resizebox{\textwidth}{!}{\input{tikzfigures/TransportThirdOrder}}
    \end{subfigure}
    \begin{subfigure}{0.49\linewidth}
        \centering
        \resizebox{\textwidth}{!}{\input{tikzfigures/TransportSecondOrder}}
    \end{subfigure}
    \caption{Convergence behavior of the 2D transport scheme. $L_2$ and $L_\infty$ error norms are plotted against the grid resolution, listed in terms of grid points $N$ and cell Reynolds number $\Re_h = hU/\nu$. In advection dominant regimes with $Re_h > 1$ the spatial convergence is third order (left), falling to second order in diffusion dominant regimes with $\Re_h < 1$ (right).}
    \label{fig:transportconvergence}
\end{figure}

The convergence of the numerical solution is measured with the $L_2$ and $L_\infty$ error norms
\begin{align}
    \epsilon_2 &= \sqrt{h^2 \sum_{\vb{x}_{ij} \notin B} \qty(\omega_{ij}(T) - \omega_{ex}(\vb{x}_{ij},T))^2}, \label{eq:l2}\\ 
    \epsilon_{\infty} &= \max_{\vb{x}_{ij} \notin B} \qty(\omega_{ij}(T) - \omega_{ex}(\vb{x}_{ij},T)). \label{eq:linf}
\end{align}
Stability constraints require that the maximum time step $\Delta t_{max}$ is $\mathcal{O}(h)$ in advection-dominant regimes ($\Re_h = (\lvert u_x \rvert + \lvert u_y \rvert )h/\nu \gg 1$) and $\mathcal{O}(h^2)$ in diffusion-dominant regimes ($\Re_h \ll 1$). Consequently, both the spatial and temporal truncation errors should be $\mathcal{O}(h^3)$ in advection-dominant regimes, and the $\mathcal{O}(h^2)$ spatial error should dominate in diffusion-dominant regimes. Figure \ref{fig:transportconvergence} plots the $L_2$ and $L_\infty$ error norms against spatial resolution for parameter sets in both regimes, demonstrating the expected rate of convergence in each one.

\section{Velocity Reconstruction}\label{section:reconstruction}

The vorticity-velocity formulation relies on a kinematic equivalence between the vorticity and velocity field.  Obtaining the vorticity from the velocity is local and inexpensive; obtaining the velocity from the vorticity, as discussed here, requires the solution of an elliptic PDE. This reconstruction procedure has been widely discussed, but rarely with a focus on 2D domains with multiple immersed bodies. In this section we review the continuous theory, and discuss the conditions under which the velocity reconstruction problem has a unique solution in multiply connected domains. This continuous formulation is then discretized using a generalization of the immersed interface Poisson solver developed by \citet{Gillis2018}, and the resulting algorithm is shown to achieve second order accuracy in reconstructing a velocity field on a multiply connected domain. 

\subsection{Continuous formulation}
Given a scalar vorticity field $\omega$ with compact support on a two-dimensional fluid domain $\Omega$, the objective of the velocity reconstruction problem is to find a divergence-free velocity field $\vb{u}$ satisfying $\nabla \times \vb{u} = \omega$, along with no through-flow boundary conditions on solid bodies. Collecting all of these requirements yields the boundary value problem
\begin{align}\label{eq:reconstruct}
\begin{split}
    \nabla \cdot \vb{u} &= 0 \qq{on} \Omega, \\
    \nabla \times \vb{u} &= \omega \qq{on} \Omega, \\
    \vu{n} \cdot \vb{u} &= \vu{n} \cdot \vb{u}_b \qq{on solid boundaries.}
\end{split}
\end{align}
In general, this is not enough to specify a unique velocity field. The question of existence and uniqueness of solutions to \eqref{eq:reconstruct} is deeply tied to the topology of $\Omega$ --- see \citet{Cantarella2002} for a complete exposition. The discussion here is limited to three cases that are common in 2D fluid simulations: unbounded domains, bounded domains, and periodic domains. 

\begin{figure}
    \centering
    \includegraphics[width=\textwidth]{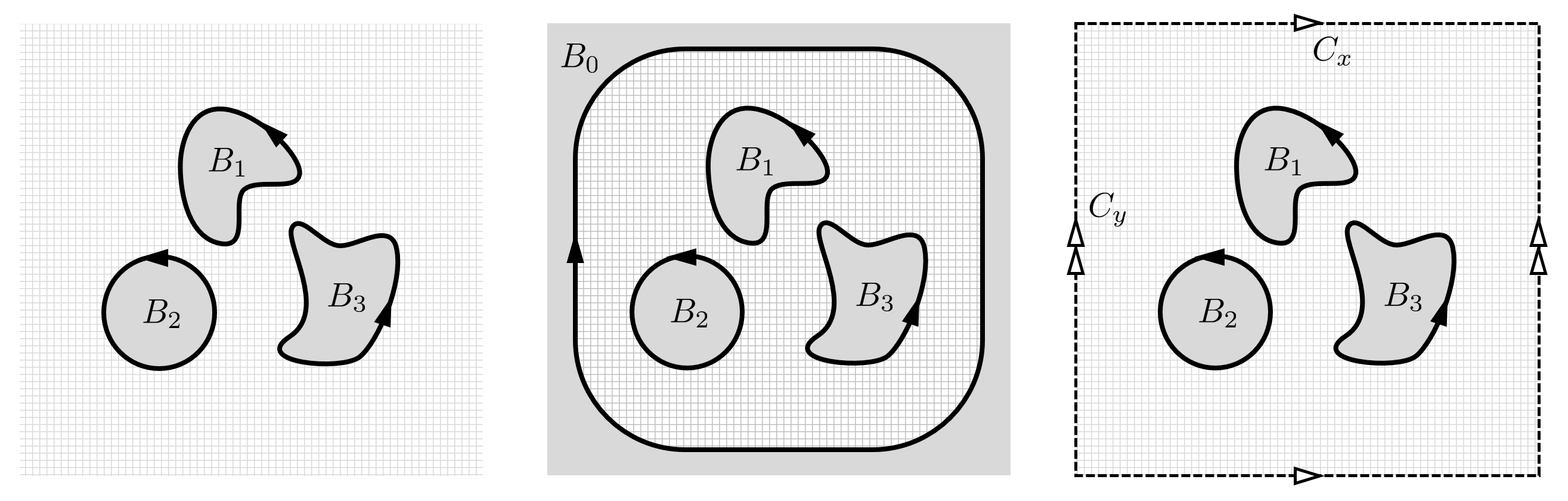}
    \caption{Three exterior boundary conditions considered for the velocity reconstruction problem: an unbounded fluid domain (left), a bounded fluid domain with solid boundaries (center), and a periodic domain with the top/bottom and left/right boundaries identified (right). Arrows indicate the direction of the tangential unit vector along each boundary.}
    \label{fig:exteriorbc}
\end{figure}

In 2D, it is simplest to analyze the reconstruction problem by re-writing the velocity field in terms of a stream function. For concreteness, assume that the fluid domain $\Omega$ is connected, that it contains distinct solid bodies $B_k$, $1 \le k \le N_b$, and that it is bounded in one of three ways: by a free-space boundary condition, by a solid exterior boundary $B_0$, or a by periodicity constraint (Figure \ref{fig:exteriorbc}). On solid boundaries, let $\vu{n}$ be a unit normal vector that points into the fluid, and define the unit tangential vector $\vu{s}$ so that $\vu{n}\times\vu{s} = \vu{k}$. A velocity field $\vb{u}(\vb{x})$ which solves \eqref{eq:reconstruct} can be written in terms of a stream function whenever $\vb{u}_b$ satisfies 
\begin{equation}\label{constarea}
    \oint_{\partial B_k} \vb{u}_b \cdot \vb{n} \dd{s} = 0 \qq{for $0 \le k \le N_b$.}
\end{equation}
This is the case whenever $\vb{u}_b$ is derived from the motion of solid bodies with constant area, and specifically holds throughout this work since we only consider stationary bodies and rotating cylinders. When \eqref{constarea} holds, let $\vb{u} = \nabla \times \psi$, so that $\nabla \cdot \vb{u} = 0$ automatically. Making this substitution in \eqref{eq:reconstruct} yields a scalar Poisson equation for the stream function, 
\begin{equation}
    -\nabla^2 \psi = \omega \qq{on} \Omega.
\end{equation}
In terms of the stream function, the no through-flow boundary condition becomes $\partial_s \psi = \vb{u}_b \cdot \vu{n}$. Under condition \eqref{constarea}, $\partial_s\psi$ can be integrated around the boundary of each body to obtain a single-valued function
 \begin{equation}
     \psi_b(s) = \int_{s_0}^s \vb{u}_b \cdot \vu{n} \dd{s}.
 \end{equation}
The no through-flow boundary condition can then be expressed as the Dirichlet boundary condition
 \begin{equation}\label{eq:psibc}
     \psi = \psi_{b} + \bar{\psi}_k \qq{on $B_k,$}
 \end{equation}
 where each $\bar{\psi}_k$ is an unknown constant associated with body $ B_k$. 
 \begin{figure}
     \centering
     \includegraphics[width=0.48\textwidth]{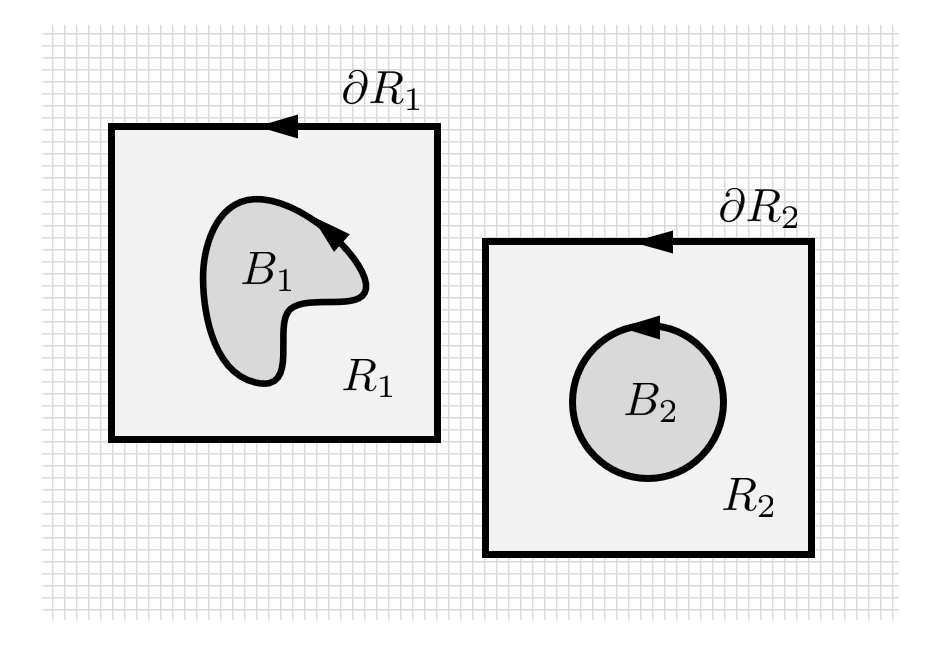}
     \caption{The circulation around the boundary of a region $R_k$ is directly related to the circulation on an enclosed object $B_k$, through the vorticity field and Stokes' theorem.}
     \label{fig:circsetup}
 \end{figure}
 
 The presence of the unknown constants $\bar{\psi}_k$ indicates that \eqref{eq:reconstruct} has multiple solutions: different choices of the constants $\bar{\psi}_k$ lead to different velocity fields, all of which are valid solutions. An effective way to single out a particular velocity field is to specify the circulation of the velocity around each solid body,
 \begin{equation}\label{eq:circconstraint}
     \oint_{\partial B_k} \vb{u} \cdot \dd{\vb{s}} = - \oint_{\partial B_k} \partial_n \psi \dd{s} = \Gamma_k \qq{for} 1 \le k \le N_b.
 \end{equation}
This provides $N_b$ scalar constraints on the stream function, which fix the $N_b$ arbitrary constants in the boundary condition. The circulation on body $B_k$ can be specified directly, or by specifying the circulation around the boundary of a region $R_k$ that contains $B_k$ (Figure \ref{fig:circsetup}). The equivalence follows immediately from Stokes' theorem, since
\begin{equation}\label{eq:circregion}
    \oint_{\partial R_k} \vb{u} \cdot \dd{\vb{s}} = \oint_{\partial B_k} \vb{u} \cdot \dd{\vb{s}} + \int_{R_k \setminus B_k} \omega \dd{A}.
\end{equation}
While this is enough to specify a unique velocity field, it still leaves some ambiguity in the stream function, which is only defined up to a global additive constant. Fixing this gauge degree of freedom, as well as enforcing an exterior boundary condition, can be handled by a method specific to each type of domain topology (Fig.~\ref{fig:exteriorbc}):
\begin{itemize}
    \item If $\Omega$ is a bounded domain, the gauge degree of freedom can be fixed by specifying a value for the arbitrary constant $\bar{\psi}_0$ on the exterior boundary $B_0$. Here this is fixed at $\bar{\psi}_0 = 0$, essentially removing this variable from the reconstruction problem and leaving a degree of freedom only on each of the $N_b$ interior solid bodies.
    \item If $\Omega$ is an unbounded domain, the stream function can be split into a free stream component $\psi_\infty = \vb{u}_\infty \times \vb{x}$ and a perturbation $\tilde{\psi}$ satisfying $\lim_{\abs{\vb{x}}\rightarrow\infty} \abs{\nabla \tilde{\psi}} = 0.$ In practice, the perturbation $\tilde{\psi}$ is calculated by a convolution between a source field and a Green's function, and the gauge degree of freedom is fixed by the choice of an arbitrary constant in the Green's function.
    \item If $\Omega$ is a rectangular domain periodic in both directions, the notion of a free stream velocity is replaced by specifying the average velocity on the horizontal periodic boundary $C_x$ and vertical periodic boundary $C_y$:
    \begin{equation}
        \bar{u}_x = \frac{1}{\abs{C_y}} \int_{C_y} \vb{u} \cdot \vu{n} \dd{s}, \quad  \bar{u}_y = \frac{1}{\abs{C_x}} \int_{C_x} \vb{u} \cdot \vu{n} \dd{s}.
    \end{equation}
    The stream function can then be split into into a periodic component $\tilde{\psi}$ and a non-periodic free stream component $\psi_\infty = \bar{\vb{u}} \times \vb{x}$. The gauge degree of freedom can be fixed by specifying that $\tilde{\psi}$ has zero mean ($\int_\Omega \tilde{\psi} \dd{A} = 0$), which is convenient for solution methods that involve a Fourier transform. Finally, in periodic domains the $N_b$ circulation constraints must satisfy the solvability condition
    \begin{equation}
        \int_\Omega \omega \dd{A} = -\sum_{i=1}^{N_b} \Gamma_k,
    \end{equation}
    which follows immediately from Stokes theorem. For $N_b = 0$ this reduces to $\int_\Omega \omega\dd{A} = 0$.
\end{itemize} 
These topology-specific exterior boundary conditions and gauge conditions, together with equations \eqref{eq:reconstruct}, \eqref{eq:psibc}, and \eqref{eq:circconstraint}, fully specify the stream function, and allow a unique velocity field to be reconstructed from the vorticity field, velocity boundary conditions, and body circulations. Here we will assume that the circulations are known in advance, and focus on discretizing the resulting elliptic equation; section \ref{section:kelvin} will address the issue of determining the body circulations.

\subsection{Immersed interface velocity reconstruction}\label{section:iimpoisson}
Because the primary component of the velocity reconstruction problem is a scalar Poisson equation, this section closely follows the 3D unbounded IIM Poisson Solver developed by \citet{Gillis2018} and applied to 2D exterior flows in \citet{Gillis2019}. The variation presented here allows for concave objects, and includes a novel discretization of the circulation constraints which allows for problems with multiple immersed bodies. 

Consider functions $\psi$ and $\omega$ defined on a Cartesian grid. The discrete Laplacian operator $\nabla_h^2$ is discretized with the standard second-order five-point finite difference stencil, so that
\begin{equation}
    \nabla_h^2 \psi_{i,j} = \frac{1}{h^2} (\psi_{i+1,j} + \psi_{i-1,j} + \psi_{i,j+1} + \psi_{i,j-1} - 4\psi_{i,j}).
\end{equation}
In a rectangular computational domain, the discretized Poisson equation $-\nabla_h^2 \psi = \omega$ can be solved efficiently with Fast Fourier Transforms (FFTs) when subject to unbounded, symmetric, or periodic boundary conditions \citep{Caprace2021}. 
This FFT-based solution procedure will be denoted by $\nabla_h^{-2}$, so that $\psi = -\nabla_h^{-2} \omega$ satisfies the discretized Poisson equation with the desired boundary treatment on the edge of the computational domain.

To extend this methodology to domains with immersed boundaries, we will consider a scalar Poisson equation $-\nabla^2 \psi = \omega$ with fixed boundary condition $\psi_b$ on solid boundaries, ignoring the unknowns $\bar{\psi}_k$ for now. We continue to view the solution $\psi$ and source field $\omega$ as functions defined on the entire Cartesian grid, now with $\psi = 0$ and $\omega = 0$ prescribed on the interior of each solid body. Thus for points that are not adjacent to the boundary, $-\nabla_h^2 \psi = \omega$ continues to hold in both the solid and fluid domains. For points in $\mathcal{A}$, the five-point Laplacian stencil crosses the solid boundary, and the IIM must be used to account for this. 

To proceed, we define some convenient notation. Consider the vector spaces
\begin{itemize}
    \item $V_G$, the space of functions defined on the entire Cartesian grid $\mathcal{G}$;
    \item $V_A \subset V_G$, the subspace of functions defined on the affected points $\mathcal{A}$;
    \item $V_C$, the space of functions defined on the control points $\mathcal{C}$.
\end{itemize}
It is also helpful to define the inclusion operator $E_a: V_A \rightarrow V_G$, which reinterprets a function with support on $\mathcal{A}$ as a function defined on all of $\mathcal{G}$ by assigning zero values to $\mathcal{G} \setminus \mathcal{A}$. In this framework, $\psi$ and $\omega$ are elements of $V_G$, while the Dirichlet boundary condition $\psi_b$ resides in $V_C$. Using this notation, a Poisson equation discretized with the IIM takes the form
\begin{equation}\label{eq:discretepoisson}
    -\nabla_h^2 \psi + E_a \gamma = \omega,
\end{equation}
where $\gamma \in V_A$ represents corrections to the standard finite difference stencil on solid boundaries. These corrections come from the polynomial extrapolation procedure outlined in section \ref{section:2diim}, here a fourth order extrapolation that uses the Dirichlet condition $\psi_b$. Consequently, $\gamma$ is a linear function of both the boundary condition $\psi_b$ and the unknown solution $\psi$, and can be written as 
\begin{equation}\label{eq:poissoncorrections}
     \gamma = A\psi + B \psi_b.
\end{equation}
Here $A:V_G \rightarrow V_A$ and $B:V_C \rightarrow V_A$ are known linear operators. 
Together equations \eqref{eq:discretepoisson} and \eqref{eq:poissoncorrections} form a system of linear equations for the unknown solution $\psi$ and the unknown IIM corrections $\gamma$. This system can be reduced via a Schur complement to a smaller system involving only $\gamma$,
\begin{equation}\label{schur}
    \qty(I_A + A\nabla_h^{-2} E_A) \gamma = -A\nabla_h^{-2} \omega + B \psi_b,
\end{equation}
where $I_A$ is the identity operator on $V_A$. The use of the solution operator $\nabla_h^{-2}$ leads to a dense linear system. However, because this solution operator can be applied efficiently using FFTs, \eqref{schur} can be solved efficiently with an iterative method. After solving for $\gamma$, the full solution $\psi$ is determined by
\begin{equation}\label{eq:psirecovery}
    \psi = -\nabla_h^{-2}(\omega - E_a \gamma).
\end{equation}
The accuracy and computational efficiency of this Schur complement approach has been demonstrated extensively \citep{Gillis2018,Gillis2019,GillisThesis}.

To extend the above methodology to the full velocity reconstruction problem, the unknown constants $\bar{\psi}_k$ must be added as additional unknowns, and the circulation constraints must be discretized to determine their values. Instead of imposing these constraints directly on immersed solid bodies, we choose to follow \eqref{eq:circregion} and specify the circulation around the boundary of a rectangular region $R_k$ that encloses each solid body $B_k$. This avoids integration over any immersed surfaces, and allows us to take advantage of a conservation property inherent in the standard 5-point Laplacian. Defining the $x$-direction numerical flux $(\nabla_h \psi)_{i+\frac 12, j} = (\psi_{i+1,j} - \psi_{i,j})/h$ and $y$-direction numerical flux $(\nabla_h \psi)_{i, j+\frac 12} = (\psi_{i,j+1} - \psi_{i,j})/h$, the operator $\nabla_h^2$ can be written as a conservative finite difference: 
\begin{align}
    \nabla_h^2 \psi &= \frac{(\nabla_h \psi)_{i+\frac 12,j} - (\nabla_h \psi)_{i-\frac 12,j}}{h} + \frac{(\nabla_h \psi)_{i,j+\frac 12} - (\nabla_h \psi)_{i,j-\frac 12}}{h}. \label{poissondiffconservation}
\end{align}
Summing \eqref{poissondiffconservation} over the points in $R_k$ yields the discrete Gauss' theorem
\begin{equation}\label{eq:poissonintconservation}
    - h^2 \sum_{R_k} \nabla_h^2 \psi = - h \sum_{\partial R_k} (\nabla_h \psi) \cdot \vu{n}.
\end{equation}
The right hand side of this relation is a second order approximation of the circulation around $\partial R_k$, which allows us to write a discrete circulation constraint
\begin{equation}\label{eq:discretecircconstraint}
    - h \sum_{\partial R_k} (\nabla_h \psi) \cdot \vu{n} = \Gamma_k.
\end{equation}
Making the substitution $\nabla_h^2 \psi = - \omega + E_A \gamma$ in \eqref{eq:poissonintconservation} leads to equivalent constraint on the unknown IIM corrections $\gamma$,
\begin{equation}
   h^2 \sum_{R_k} \omega - h^2 \sum_{R_k} E_A \gamma = \Gamma_k.
\end{equation}
To incorporate this constraint into an immersed interface Poisson solver, consider the following notation:
\begin{itemize}
    \item $\sum_{B_k}: V_A \rightarrow \mathbb{R}$ sums all function values at affected points associated with $B_k$. 
    \item $\sum_{R_k} : V_G \rightarrow \mathbb{R}$ sums all function values which fall within a region $R_k$.
    \item $\mathbb{1}_{k} \in V_C$ is the vector with ones at control points associated with $B_k$ and zeros otherwise.
\end{itemize}
Introducing the circulation constraints as additional equations and the boundary constants $\bar{\psi}_k$ as unknowns, the discretized reconstruction problem becomes
\begin{align}
    \qty(I_A + A\nabla_h^{-2} E_a)\gamma - B\sum_{k=1}^{N_b} \bar{\psi}_k \mathbb{1}_k &= - A\nabla_h^{-2} \omega + B \psi_b, \\
    \Sigma_{B_k} \gamma &= - \Gamma_{R_k} + \Sigma_{R_k} \omega \qq{for} 1 \le k\le N_b.
\end{align}
These equations are solved iteratively with the GMRes algorithm to find the values of the unknowns $(\gamma, \; \bar{\psi}_k)$, and the full solution $\psi$ is recovered using \eqref{eq:psirecovery}.

Once the stream function has been determined, the velocity field $\vb{u} = \nabla \times \psi$ can be recovered. To do so in the presence of immersed bodies, the stream function is extrapolated beyond the solid boundaries using the boundary condition $\psi_b + \bar{\psi}_k$. The extrapolation used here is the same fourth-order procedure calculated by the operators $A$ and $B$ in the discretized reconstruction problem. The velocity at grid points within the domain is then calculated with the second-order centered difference
\begin{equation}
    \vb{u}_{i,j} = \qty(\frac{\psi_{i,j+1} - \psi_{i,j-1}}{2h}, -\frac{\psi_{i+1,j} - \psi_{i-1,j}}{2h})^\text{T}.
\end{equation}
 While a third-order extrapolation would be sufficient to maintain second order accuracy in the velocity field, we observed that the fourth-order extrapolation leads to a significantly reduced truncation error in the velocity field near solid boundaries, where velocity gradients tend to be the largest.

\subsection{Numerical results}\label{section:reconstructionconvergence}

\begin{figure}
    \centering
    \begin{subfigure}{0.49\linewidth}
        \centering
        \includegraphics[width=0.72\textwidth]{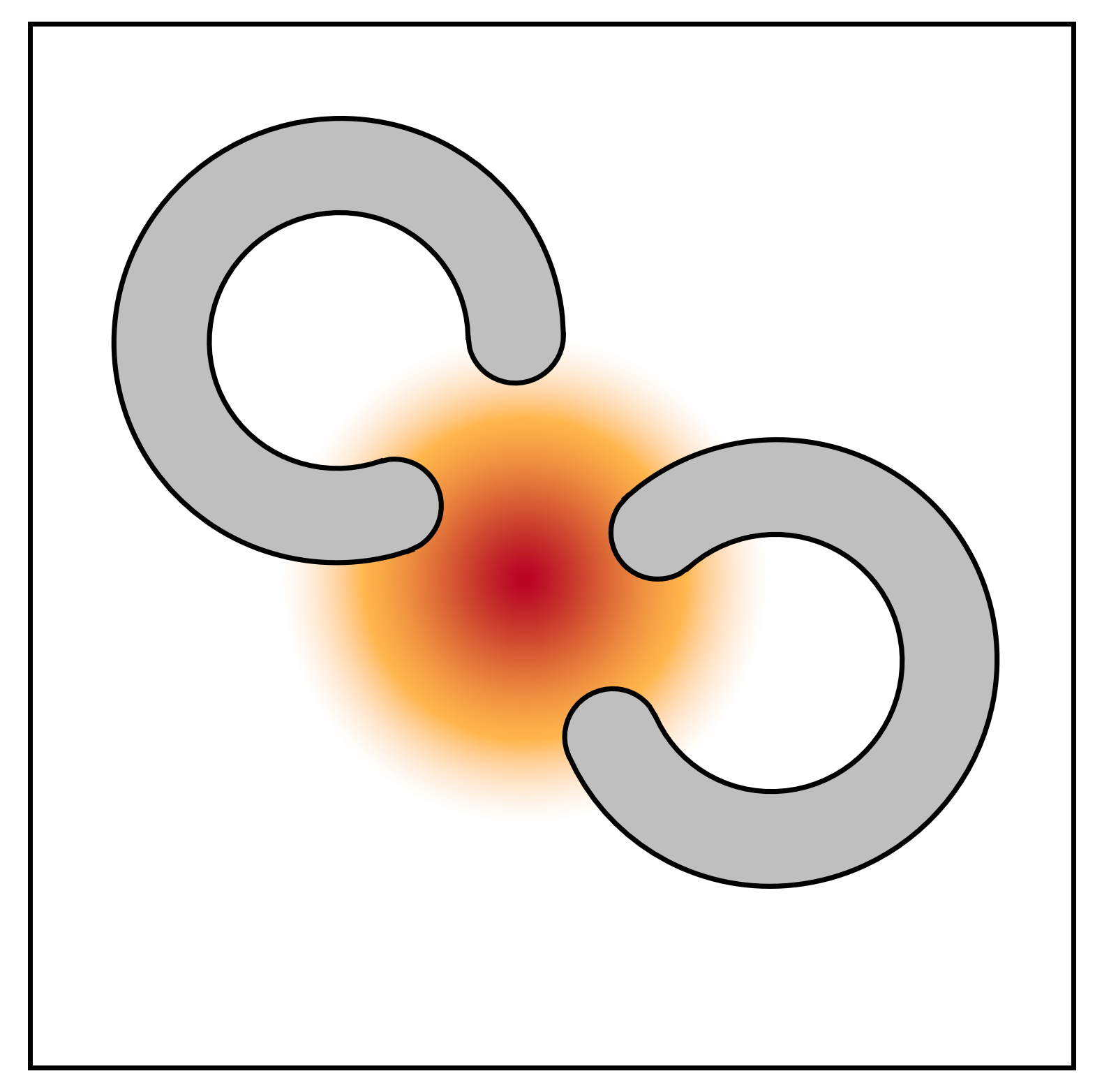}
        \caption{}
        \label{fig:poissonparameters}
    \end{subfigure}
    \begin{subfigure}{0.49\linewidth}
        \centering
        \resizebox{\textwidth}{!}{\input{tikzfigures/LambOseenConcave}}
        \caption{}
        \label{fig:poissonconvergence}
    \end{subfigure}
    \par\bigskip
    \begin{subfigure}{\linewidth}
        \centering
        \begin{tabular}{ | c | c | } 
        \hline
        Element & Parameters \\ 
        \hline\hline
        Grid & $\vb{x}_{ij} = (i/N, \; j/N)$ for $0 \le i,j \le N - 1$. \\
        \hline
        Vortex & $\vb{x}_0 = (0.501,0.501), \Gamma = 1.0 , \nu t = 0.0025$ \\
        \hline 
        Body 1 & $\vb{x}_c = (0.309,0.679), r_a=0.167 , r_b =0.057 , \theta_a = 0.5, \theta_b = 1.8$ \\
        \hline
        Body 2 & $\vb{x}_c = (0.559,0.451), r_a=0.157 , r_b=0.057 , \theta_a = 3.7 , \theta_b = 2.0$ \\
        \hline
        \end{tabular}
        \caption{}
        \label{tab:reconstruction}
    \end{subfigure}
    \caption{(a and c) Geometry and parameter values used in the convergence study. (b) $L_2$ and $L_\infty$ error norms for the velocity field as a function of grid resolution, showing second order convergence in both. }
\end{figure}

To demonstrate the accuracy and flexibility of our velocity reconstruction procedure, we consider a test case with non-convex solid bodies, a multiply connected domain, and free-space boundary conditions (Figure \ref{fig:poissonparameters}). Two bodies with geometry described in Figure \ref{fig:transportsetup} are superimposed on the vorticity field of a Lamb-Oseen vortex, 
\begin{equation}\label{eq:lovorticity}
    \omega_{LO}(\vb{x}) = \frac{\Gamma}{4\pi\nu t} \exp(-\frac{\norm{\vb{x} - \vb{x}_0}^2}{4\nu t}),
\end{equation}
and the velocity $\vb{u}_b$ on their boundaries is prescribed to match the corresponding velocity field
\begin{equation}\label{eq:lovelocity}
    \vb{u}_{LO}(\vb{x}) = \frac{\Gamma}{2\pi r^2}\qty(1 -  \exp(-\frac{\norm{\vb{x} - \vb{x}_0}^2}{4\nu t})) \vu{k} \times (\vb{x} - \vb{x}_0).
\end{equation}
The reconstruction procedure described in the previous section is then used to recover the full velocity field, which is compared to the exact solution $\vb{u}_{LO}(\vb{x})$. This is done on a square Cartesian grid with $N$ points along each side. To prescribe the circulations on solid bodies, each body is enclosed in a bounding region $R_k$, and the circulation on $\partial R_k$ is estimated to second order from the original vorticity field:
\begin{equation}
    \Gamma_{R_k} = h^2 \sum_{\vb{x}_{ij} \in R_k} \omega_{LO}(\vb{x}_{ij}).
\end{equation}
The full details of the geometry, flow parameters, and bounding regions are provided in Table \ref{tab:reconstruction}. Figure \ref{fig:poissonconvergence} plots the $L_2$ and $L_\infty$ norms of the velocity error $\norm{\vb{u} - \vb{u}_{exact}}$ against the spatial resolution $N$, demonstrating second order convergence in both error norms.  

\section{Navier-Stokes}\label{section:navierstokes}

In this section we introduce the components of our full Navier-Stokes discretization that do not fit neatly within the vorticity transport or velocity reconstruction problems. This includes a method for enforcing Kelvin's theorem, which is key to simulations with multiple bodies; an outflow boundary condition for external flows; the vorticity boundary condition for the transport equation; and the calculation of forces and surface tractions. These are then combined with the transport scheme developed in section \ref{section:transport} and reconstruction scheme developed in section \ref{section:reconstruction} to create an algorithm that solves the full Navier-Stokes equations. 

\subsection{Circulation in multiply connected domains}\label{section:kelvin}
Section \ref{section:reconstruction} outlined an algorithm that reconstructs the velocity field from the vorticity field, provided that the circulation around each solid body is known in advance. Continuously, the circulation around any body satisfying a no-slip condition can be determined by integrating the tangential component of the prescribed boundary velocity. However, for numerical algorithms that enforces the no-slip condition only approximately, including ours, a different strategy is needed to specify these circulations. Here we outline a method based on the enforcement of a discrete form of Kelvin's theorem.

For a 2D viscous flow, Kelvin's theorem states that the circulation around any material contour $C(t)$ evolves according to
\begin{equation}\label{materialKelvin}
    \dv{}{t} \oint_{C(t)} \vb{u} \cdot \dd{\vb{s}} = -\nu \oint_{C(t)} \pdv{\omega}{n} \dd{s}.
\end{equation}
If $C$ is a stationary contour, then an application of Reynolds transport theorem gives the equivalent expression
\begin{equation}\label{Kelvin}
    \dv{}{t} \oint_{C} \vb{u} \cdot \dd{\vb{s}} = - \oint_{C} \qty(\vb{u} \omega - \nu \nabla \omega) \cdot \vu{n} \dd{s},
\end{equation}
which is an ordinary differential equation that governs the evolution of the circulation around $C$. If $C$ is the boundary of a simply connected fluid region $R$, then (\ref{Kelvin}) can be derived directly from the vorticity transport equation and Stokes' theorem. However, if $C$ is not the boundary of a simply connected fluid region, then Kelvin's theorem is a separate constraint from the vorticity transport equation that ensures the existence of single-valued pressure field \citep{Lequeurre2020}.

For immersed interface methods, enforcing Kelvin's theorem has been challenging. In \citep{MarichalThesis}, Marichal observes that prescribing $\Gamma_{\partial B} = \Sigma_B \gamma_h = 0$ on a stationary solid body during velocity reconstruction leads to an unstable numerical scheme. The total circulation of the flow --- as measured by the sum of all vorticity values on the grid --- increases rapidly with time. The author avoids this instability by instead prescribing zero far-field circulation, through the condition
\begin{equation}\label{far-field}
    \Gamma_{\partial B} + h^2 \sum_{R_\infty} \omega = 0. 
\end{equation}
Here the vorticity field is defined to be zero inside solid bodies, $R_\infty$ is a rectangular grid-aligned region containing the full support of the vorticity field, and we use the summation notation defined in equation~\eqref{eq:sum_R_def}. This condition directly enforces Kelvin's theorem by linking the circulation around a solid boundary to the  vorticity created on that boundary. However, it does not generalize to flows in which vorticity is allowed to leave the computational domain at outflow boundaries, and it can only uniquely determine the circulation of one immersed body. 

To generalize this condition, we note here that when both the vorticity transport equation and the velocity reconstruction problem are discretized with conservative finite differences, as we have done in sections \ref{section:transport} and \ref{section:reconstruction}, a discrete form of Kelvin's theorem holds automatically on the boundary of any rectangular grid-aligned fluid region. For any such region $R$, combining the conservation property of the transport equation \eqref{2dintconserve} with the discrete Stokes theorem enforced by the reconstruction procedure \eqref{eq:poissonintconservation} leads to the expression
\begin{equation}\label{discreteKelvin}
    -h \dv{}{t} \sum_{\partial R_k} (\nabla_h \psi) \cdot \vu{n} = -h \sum_{\partial R} (\vb{u}\omega - \nu \nabla_h\omega) \cdot \vu{n},
\end{equation}
which is analogous to \eqref{Kelvin}. This discrete form of Kelvin's theorem immediately extends to any region which can be built by adding or removing a series of grid-aligned rectangles, each satisfying \eqref{discreteKelvin}. If the region $R$ is not purely fluid, but encloses a single immersed body $B$, then \eqref{discreteKelvin} no longer holds automatically. To remedy this, we begin by selecting $R$ as the bounding region enclosing $B$ in the velocity reconstruction procedure. We then treat Kelvin's theorem as an ODE which defines the evolution of the circulation $\Gamma_R$ associated with $R$:
\begin{equation}\label{discreteKelvinChoice}
    \dv{\Gamma_R}{t} = -h \sum_{\partial R} (\vb{u}\omega - \nu \nabla_h\omega) \cdot \vu{n}.
\end{equation}
This ODE is integrated with the same Runge-Kutta method as used in the transport equation, so that \eqref{discreteKelvin} holds automatically on $\partial R$. Condition \eqref{discreteKelvinChoice} also enforces a discrete Kelvin's theorem on the boundary of every other grid-aligned region enclosing $B$, since any such region can be built from  $R$ by adding or removing rectangular fluid regions satisfying \eqref{discreteKelvin}. Thus any choice of the region $R$ on which we enforce \eqref{discreteKelvinChoice} is equivalent to any other, and the dynamics of the discretized fluid system are independent of our choice of bounding regions. 

Taking $R$ to be a contour $R_\infty$ which encloses the entire vorticity field, we immediately find that condition \eqref{discreteKelvinChoice} is equivalent to the far-field condition \eqref{far-field} proposed by Marichal for flows with a single immersed body. However, condition \eqref{discreteKelvinChoice} does not require a contour which encloses the entire vorticity field, allowing for simulations with outflow boundary conditions. It can be also generalized to flows with multiple bodies by choosing a separate bounding region $R_k$ for each body $B_k$. Here again, the specific choice of each bounding region $R_k$ does not affect the discrete dynamics. 

\subsection{Outflow boundaries}\label{section:outflow}

The previous section introduced a circulation tracking scheme for vorticity-based immersed interface methods that does not require the computational domain to contain the entire vorticity field. This allows for the use of outflow boundary conditions, which are essential for long-time simulations of external flows. The outflow condition used here is a 2D analogue of the condition used for simulations of 3D wake dynamics in \citep{GillisThesis, Chatelain2013, Caprace2020}. For a purely horizontal free stream, a vertical outflow plane is specified downstream of all immersed bodies, and the vorticity field is mirrored across this outflow plane. This leads to a reconstructed velocity field that satisfies
\begin{align}
    \begin{split}
        \pdv{u_x}{x} &= 0, \\
        u_y &= 0
    \end{split}
\end{align}
on the outflow plane. The portion of the mirrored vorticity field adjacent to the outflow plane is used when calculating the vorticity flux in the vorticity transport equation. The rest of this mirror distribution is never constructed explicitly; it enters the velocity reconstruction problem through the use of a symmetry boundary condition in the operator $\nabla_h^{-2}$, as described by \citet{Caprace2021}.

\begin{figure}
    \centering
    \includegraphics[width=\textwidth]{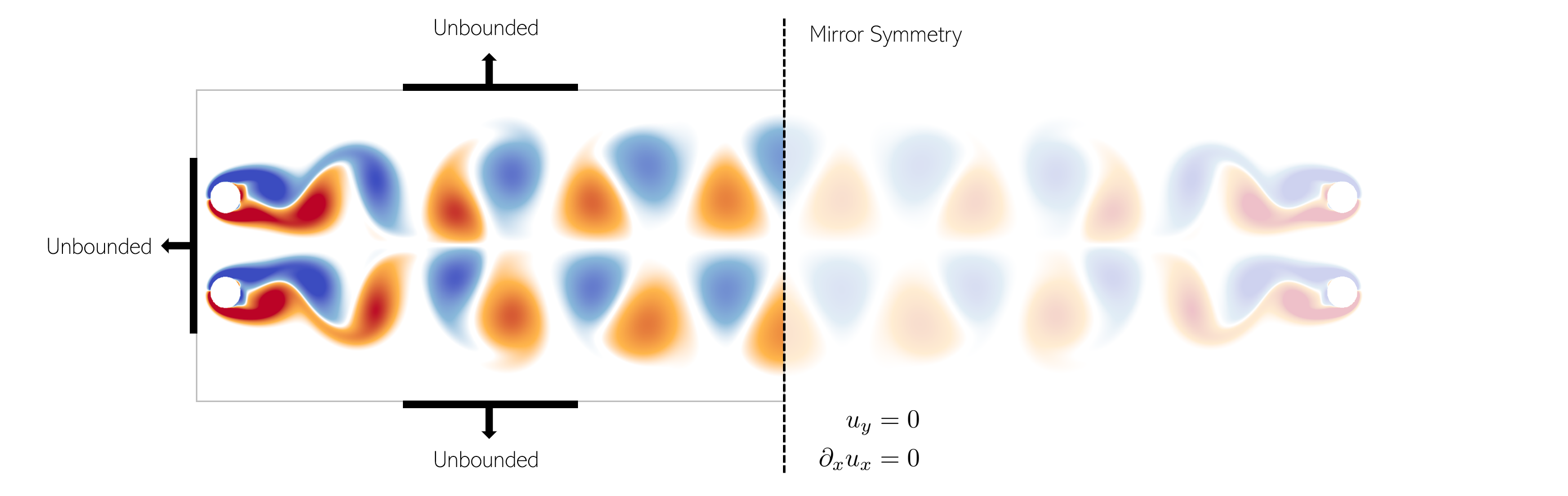}
    \caption{Outflow boundaries are approximated by an even symmetry condition for the vorticity field on the outflow plane.}
    \label{fig:outflow}
\end{figure}

\subsection{Vorticity boundary conditions and vorticity flux}\label{section:vorticitybc}
The vorticity-velocity form of the Navier-Stokes equations requires a no-slip velocity boundary condition on immersed surfaces. However, because the velocity field is reconstructed from the vorticity field through an elliptic equation, it is difficult to translate the no-slip velocity boundary condition into a boundary condition for the vorticity transport equation. The approach taken here is a minor variation on the method used by \citet{Gillis2019}, which allows for high order explicit time integration and nonconvex immersed bodies. It is similar in spirit to the immersed interface vorticity boundary condition used by \citet{Linnick2005}, and falls into the class of local vorticity boundary conditions originated by \citet{Thom1933} and catalogued by \citet{E1996}. Other notable strategies are the vorticity integral constraints developed by \citet{Quartapelle1993}, and the Lighthill splitting approach (which was investigated extensively in the context of immersed interface methods by \citet{MarichalThesis}). However, neither the integral constraints nor Lighthill splitting allow for the use of a high order explicit time integration scheme; Quartapelle's integral constraints require implicit integration, while Lighthill's splitting method is limited to first order temporal accuracy. 

\begin{figure}
    \centering
    \includegraphics[width=0.55\textwidth]{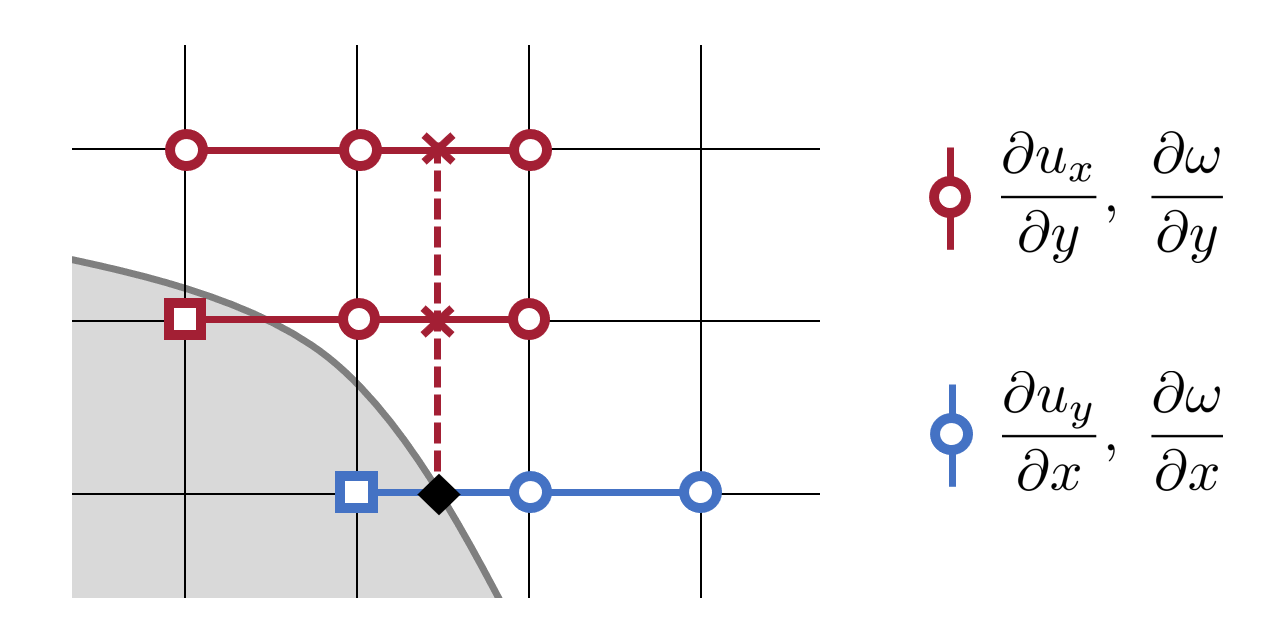}
    \caption{Stencils used to calculate the vorticity boundary condition at second order. The blue points define a three-point stencil for the x-direction velocity derivative on the boundary, which uses three neighboring velocity values defined on the Cartesian grid. The y-direction velocity derivative is more involved; the value of $u_x$ at the red crosses is interpolated using three-point stencils, and these values are then used along with the velocity boundary condition to calculate the y-direction velocity derivative at the boundary. This same set of stencils is also used to calculate the boundary vorticity gradient.}
    \label{fig:vorticitybc}
\end{figure}

In the current method, the velocity reconstruction process yields a velocity field $\vb{u}$ that is defined on the Cartesian grid points. The velocity field at control points $\vb{x}_c \in \mathcal{C}$ is known from the velocity boundary condition $\vb{u}_b$. Using this information, both components of the velocity field are extended past the domain boundary using a third order polynomial extrapolation. The boundary vorticity $\omega_b$ is taken to be the curl of the extended velocity field evaluated on the boundary, and is evaluated using the second-order stencils shown in Figure \ref{fig:vorticitybc}. This is a more compact scheme than the stencils used by \citet{Gillis2019}, intended to better accommodate concave geometries.

\subsection{Force calculations}\label{section:forces}

The total lift force and drag force acting on an immersed body can be calculated using the control volume formulations derived by \citet{Noca1997}, and the total torque acting the body can be calculated using an analogous control volume formulation which we derive in \ref{appendix:forces}. The shear stress distribution on a stationary immersed surface can be derived directly from the surface vorticity: $\tau(s) = \nu \omega(s)$, where $s$ is a surface coordinate. 

Recovering the surface pressure distribution is more involved, and relies on the relation between the surface pressure gradient and surface vorticity gradient. On stationary solid boundaries with a no-slip condition, as considered here, this is
\begin{equation}\label{pressuregrad1}
     0 = - \nabla p - \nu \nabla \times \omega.
\end{equation}
To evaluate the vorticity gradient, the vorticity field is extended past the domain boundary using a third order extrapolation, taking into account the vorticity boundary condition $\omega_b$ computed as in section \ref{section:vorticitybc}. The derivative of this extended field along the coordinate directions is calculated using the same stencils points as the vorticity boundary condition (Figure \ref{fig:vorticitybc}), and then projected onto the local normal and tangential unit vectors. With this gradient known, we can consider two distinct methods of pressure recovery. In the first, the tangential pressure gradient is calculated from the normal vorticity flux, and then integrated over the immersed surface. Beginning the integration at an arbitrary point with surface coordinate $s_0$ leads to the expression
\begin{equation}\label{pressureflux}
    p(s) - p(s_0) = \int_{s_0}^s \nu \pdv{\omega}{n} \dd{s}.
\end{equation}
Because the fluid adjacent to the boundary forms a material contour, Kelvin's theorem \eqref{materialKelvin} guarantees that this integral is single-valued. The additive constant $p(s_0)$ cannot be determined from boundary information alone, but this procedure is still useful for measuring relative pressure differences over the immersed surface. To avoid the unknown additive constant $p(s_0)$, we also consider a second and more expensive method of pressure recovery. Following \citet{Lee2014}, we define the total pressure
\begin{equation}
    H = (p - p_\infty) + \frac{1}{2} \qty(\abs{\vb{u}}^2 - \abs{\vb{u}_\infty}^2),
\end{equation}
which satisfies the far-field condition $H \rightarrow 0$ as $x \rightarrow \infty$ and the scalar Poisson equation
\begin{equation}\label{eq:pressurepoisson}
    \nabla^2 H = \nabla \cdot (\vb{u} \times \omega\vu{k}).
\end{equation}
A Neumann boundary condition for the equation can be constructed from the definition of $H$ and the normal component of (\ref{pressuregrad1}):
\begin{equation}\label{eq:totalpressurebc}
    \eval{\pdv{H}{n}}_S = \pdv{p}{n} - \pdv{}{n}\qty(\frac{\abs{\vb{u}}^2}{2}) = - \nu \pdv{\omega}{s} - \pdv{}{n}\qty(\frac{\abs{\vb{u}}^2}{2}).
\end{equation}
Discretely, the tangential vorticity gradient and the normal gradient of $\abs{\vb{u}}^2/2$ are calculated using the boundary stencils described in section \ref{section:vorticitybc}. The pressure Poisson equation \eqref{eq:pressurepoisson} is then solved using the IIM Poisson solver outlined in section \ref{section:iimpoisson}. To handle the Neumann boundary condition, we use the compatible extrapolation procedure developed by \citet{Marichal2016}. The resulting pressure field is interpolated back to the immersed solid boundaries using a third order polynomial extrapolation to obtain the surface pressure $p(s)$.

\subsection{Full algorithm for flow simulations}
Having established a transport scheme, a velocity reconstruction scheme, a method for enforcing Kelvin's theorem, and a vorticity boundary condition, we can lay out a complete algorithm for solving the 2D incompressible Navier-Stokes equations in vorticity form. The dynamics variables are the discretized vorticity field $\omega(\vb{x}, t)$ and the $N_b$ bounding box circulations $\Gamma_k(t)$, both of which require an initial conditions $\omega_0(\vb{x})$ and $\qty{\Gamma_{k,0} }$. These can be specified directly, or inferred from an initial velocity field $\vb{u}_0(\vb{x})$. Together the discretized vorticity field and circulation are determined by a large system of ODEs,
\begin{equation}
    \pdv{\omega}{t}, \; \qty{\dv{\Gamma_k}{t}} = f\qty(\omega, \qty{\Gamma_k}),
\end{equation}
defined by the following sequence:

\begin{itemize}
    \item \textbf{Velocity Reconstruction.} Given the vorticity field $\omega$, circulations $\qty{\Gamma_k}$, and velocity boundary condition $\vb{u}_b$, the stream function $\psi$ is recovered by solving the scalar Poisson equation
    \begin{align*}
         -\nabla^2 \psi &= \omega  \qq{on} \Omega, \\[0.5ex]
         \psi &= \psi_b + \bar{\psi}_k \qq{on} \partial B_k,\\
         -\oint_{R_k} \partial_n \psi \dd{s} &= \Gamma_k \qq{for} 1 \le k \le N_b.
    \end{align*}
    This is done with the reconstruction procedure outlined in section \ref{section:reconstruction}. The velocity field $\vb{u}$ is calculated by differentiating the stream function.
    \item \textbf{Vorticity Transport.} The Dirichlet boundary condition $\omega_b$ for the vorticity transport equation is calculated using the local method outlined in section \ref{section:vorticitybc}. The transport equation 
    \begin{equation*}
        \pdv{\omega}{t} =  -\nabla \cdot (\vb{u}\omega - \nu \nabla \omega)
    \end{equation*} 
    then determines the time derivative of the vorticity field, through the conservative spatial discretization developed in section \ref{section:transport}.
    \item \textbf{Kelvin's Theorem.} As outlined in section \ref{section:kelvin}, the time derivative of the circulations is determined by Kelvin's theorem, 
    \begin{equation*}
        \dv{\Gamma_k}{t} = -\oint_{\partial R_k} (\vb{u}\omega - \nu \nabla \omega).
    \end{equation*}
    The spatial integration is performed with the same vorticity flux used in the transport scheme. 
\end{itemize}
This system of ODEs is integrated in time with a low storage third-order Runge-Kutta scheme \citep{Williamson1980}. The size of each time step is chosen to be a fixed fraction $0 < C_{\mathrm{stab}} < 1$ of the maximum stable time step for the transport scheme, calculated using the procedure outlined in \ref{section:transportstability}. All force and pressure calculations are performed after the first velocity reconstruction of each time step, when the vorticity and circulations have third-order temporal accuracy. 

The algorithm described above has been implemented in C++ using Cubism \citep{Hejazialhosseini2012}, a library for block-based parallelism on uniform resolution Cartesian grids. The velocity reconstruction problem is solved with FFT-accelerated convolutions performed by FLUPS \citep{Caprace2021}, which allows for fast solutions of scalar Poisson equations on rectangular domains with arbitrary combinations of unbounded, symmetric, and periodic boundary conditions.

\section{Results}\label{section:results}
The Navier-Stokes discretization developed in the previous sections is applicable to a broad class of 2D incompressible flows. Here we first demonstrate the convergence of the method for a simple test case with an analytical solution, and then illustrate the effectiveness of this discretization in calculating vorticity fields, velocity fields, and surface traction distributions for a variety of external and internal flows.

\subsection{Convergence: Lamb-Oseen vortex}
To demonstrate the convergence of the 2D Navier-Stokes discretization developed here, we consider an external flow test case with a solid body and an analytical solution, as done in \citet{Gillis2019}. A rotating cylinder with radius $R$ and center $\vb{x}_c$ is immersed in a uniform Cartesian grid with grid spacing $h$. The initial vorticity field outside of cylinder is set to match the vorticity field of a Lamb-Oseen vortex centered at $\vb{x}_c$, and the cylinder's time-dependent rotation rate is set so that at each point on the solid boundary $\vb{u}_b(t) = \vb{u}_{LO}(R, t)$. This ensures that the Lamb-Oseen vortex is an analytical solution to the flow outside of the cylinder which satisfies the no-slip boundary condition for all time. The flow is integrated from time $t_0$ to time $t_f$ using a third-order Runge-Kutta scheme, and the numerical vorticity and velocity fields are compared to the analytical vorticity and velocity fields using the $L_2$ and $L_\infty$ error norms defined in sections \ref{section:transportconvergence} and \ref{section:reconstructionconvergence}. The full details of the grid, cylinder, vortex, and time integration are provided in Table \ref{tab:nsconvergence}.

Figure \ref{fig:nsconvergence} plots the $L_2$ and $L_\infty$ error norms of the velocity and vorticity fields against the spatial resolution $h$, demonstrating second order convergence in both norms for both fields. We emphasize that the convergence rate of the $L_\infty$ error norms indicates that the full Navier-Stokes algorithm achieves second-order spatial accuracy right up to the immersed boundary. 
\begin{figure}[htb]
    \centering
    \begin{subfigure}{0.49\textwidth}
        \centering
        \resizebox{\textwidth}{!}{\input{tikzfigures/NSVorticity}}
        \caption{Vorticity error.}
    \end{subfigure}
    \begin{subfigure}{0.49\linewidth}
        \centering
        \resizebox{\textwidth}{!}{\input{tikzfigures/NSVelocity}}
        \caption{Velocity Error}
    \end{subfigure}
    \par\bigskip
    \begin{subfigure}{\linewidth}
    \centering
    \begin{tabular}{ | c | c | } 
    \hline
    Element & Parameters \\ 
    \hline\hline
    Grid & $\vb{x}_{ij} = (i/N, \; j/N)$ for $0 \le i,j \le N - 1$. \\
    \hline
    Time & RK3 with safety factor $C_{stab} = 0.7$. $t_0 = 1.0$, $t_f = 2.0$. \\
    \hline
    Cylinder & $\vb{x}_c = (0.507,0.507)$, $R = 0.15$ \\
    \hline
    Vortex & $\Gamma = \pi$ , $\nu = 0.001$ \\
    \hline 
    \end{tabular}
    \caption{Parameters used in the full Navier-Stokes convergence study.}
    \label{tab:nsconvergence}
    \end{subfigure}
    \caption{Convergence of the error in the (a) vorticity field and (b) velocity field with spatial resolution for the full Navier-Stokes discretization, as well as the parameter values used in the convergence study (c).}
    \label{fig:nsconvergence}
\end{figure}

\subsection{Impulsively rotated cylinder}
The flow around an impulsively rotated cylinder is another viscous exterior flow with an analytical solution, and an excellent test case for the enforcement of Kelvin's theorem. Consider a cylinder of radius $R$ immersed in a quiescent fluid with viscosity $\nu$, which begins rotating with constant angular velocity $\Omega$ at time $t = 0$. The impulsive start releases a singular vortex sheet into the flow, which then diffuses radially outward. Any mismatch between the magnitude of this thin vortex sheet and the rotation rate of the object leads to a violation of Kelvin's theorem, and can cause significant errors in the vorticity field and the resulting viscous moment acting on the cylinder.

For simplicity, the flow around the cylinder is assumed to remain axisymmetric. Using a non-dimensional time $t^* = \nu t/R^2$ and radial coordinate $r^* = r/R$, the non-dimensional vorticity distribution $\omega^* = \omega/\Omega$ is given by
\begin{equation}\label{besselw}
    \omega^*(r^*, t^*) = -\frac{2}{\pi}\int_0^\infty\real\qty{\frac{K_0(ir^*x)}{K_1(ix)}}e^{-x^2t^*}\dd{x},
\end{equation}
while the non-dimensional velocity distribution $u^* = u_\theta/R\Omega$ is given by
\begin{equation}\label{besselu}
    u^*(r^*, t^*) = \frac{1}{r^*}-\frac{2}{\pi}\int_0^\infty\imaginary\qty{\frac{K_1(ir^*x)}{K_1(ix)}}\frac{e^{-x^2t^*}}{x}\dd{x}.
\end{equation}
Here $K_0(x)$ and $K_1(x)$ are modified Bessel functions of the second kind, while $\real$ and $\imaginary$ denote the real and imaginary parts respectively. These analytical expressions are derived in \ref{section:improt} and provide a more easily evaluated result for the velocity field compared to the expressions provided by \citet{Lagerstrom1996}. The integrands in (\ref{besselw}) and (\ref{besselu}) are non-singular at $x = 0$ and decay rapidly as $x \rightarrow \infty$, so that the improper integrals can be evaluated numerically with good accuracy. The total non-dimensional moment $M^* = M/2\pi R^2\nu\Omega$ acting on the cylinder can be calculated by integrating the resulting shear stress distribution, giving $M^*(t^*) = \omega^*(1, t^*) - 2$. 

\begin{figure}
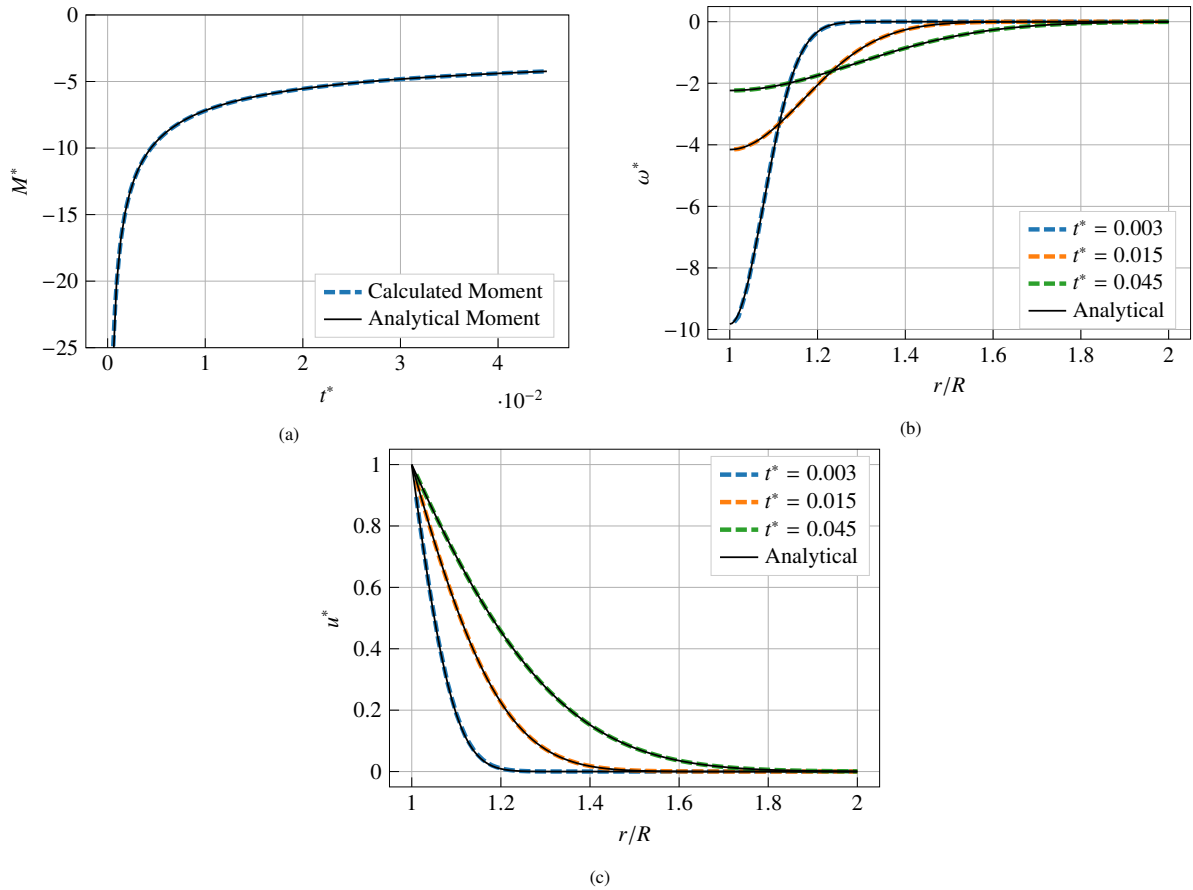

    \centering
    \begin{subfigure}{0.46\linewidth}
        \centering
        \resizebox{\textwidth}{!}{\input{tikzfigures/RotationMoment8x8}}
        \caption{}
    \end{subfigure}
    \hspace{0.03\textwidth}
    \begin{subfigure}{0.46\textwidth}
        \centering
        \resizebox{\textwidth}{!}{\input{tikzfigures/RotationVorticity8x8}}
        \caption{}
    \end{subfigure}
    \vspace{10pt}
    \begin{subfigure}{0.46\linewidth}
        \centering
        \resizebox{\textwidth}{!}{\input{tikzfigures/RotationVelocity8x8}}
        \caption{}
    \end{subfigure}
    \caption{Numerical and semi-analytical results for the impulsively rotated cylinder, including (a) the time history of the total moment, (b) the vorticity field at selected times, and (c) the velocity field at selected times.}
    \label{fig:cylmom}
\end{figure}

These analytical results are independent of the Reynolds number $\mathrm{Re}_\Omega = R^2\Omega/\nu$. For the simulations discussed here, we have chosen $\Re _\Omega = 50$ to avoid any high Reynolds number instabilities that may disrupt the axisymmetric flow. Figure \ref{fig:cylmom} shows the time evolution of the moment acting on the impulsively rotated cylinder for $t^* \le 0.045$, along with velocity and vorticity profiles at selected times. At the resolution shown here ($D/h = 104.2$), the numerical and semi-analytical results are in excellent agreement. 

\subsection{Impulsively translated cylinder} \label{section:impcylresults}
The impulsively translated cylinder is a widely used test case in two-dimensional incompressible flow \citep{ Koumoutsakos1995, Lee2014, Anderson1996, Gillis2019, Wu2019}. Consider a cylinder of diameter $D$ and center $\vb{x}_c = \qty(x_c, \ y_c)$ immersed in an unbounded fluid with kinematic viscosity $\nu$. At time $t = 0$, the cylinder begins translating with constant velocity, which produces a free-stream velocity of $\vb{u}_\infty = (u_{\infty,x}, \ u_{\infty,y})$ in a reference frame attached to the cylinder. The dynamics of the resulting flow depend only on the Reynolds number $\mathrm{Re}_D = u_\infty D / \nu$. This section focuses on the short-time evolution of the flow-field, which takes place before the symmetry of the problem is broken and the commonly-observed vortex shedding behavior begins. In this symmetric regime there is no lift and no net moment acting on the cylinder, so that the drag force is the only relevant load. 

\begin{figure}
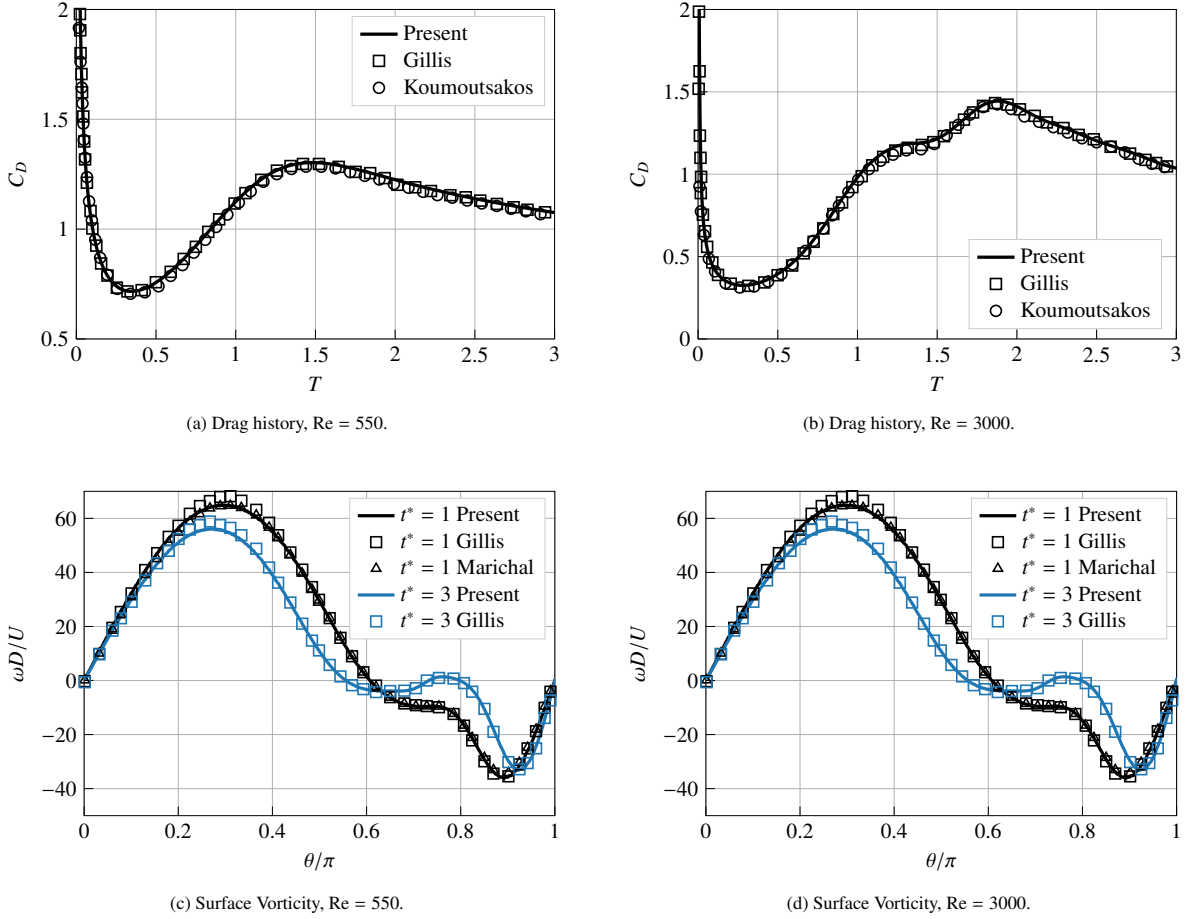

    \centering
    \begin{subfigure}{0.46\linewidth}
        \centering
        \resizebox{\textwidth}{!}{\input{tikzfigures/Cyl_Re550_Drag}}
        \caption{Drag history, $\Re = 550$.}
    \end{subfigure}
    \hspace{0.03\textwidth}
    \begin{subfigure}{0.46\linewidth}
        \centering
        \resizebox{\textwidth}{!}{\input{tikzfigures/Cyl_Re3000_Drag}}
        \caption{Drag history, $\Re = 3000$.}
    \end{subfigure}
    \par\vspace{20pt}    
    \begin{subfigure}{0.46\linewidth}
        \centering
        \resizebox{\textwidth}{!}{\input{tikzfigures/Cyl_Re550_Vort}}
        \caption{Surface Vorticity, $\Re = 550$.}
    \end{subfigure}
    \hspace{0.03\textwidth}
    \begin{subfigure}{0.46\linewidth}
        \centering
        \resizebox{\textwidth}{!}{\input{tikzfigures/Cyl_Re3000_Vort}}
        \caption{Surface Vorticity, $\Re = 3000$.}
    \end{subfigure}
    \caption{Drag history and surface vorticity profiles for the impulsively started cylinder at $\Re = 550$, $N_\delta = 8.73$, $Q = 0.31$ (left column) and $Re = 3000$, $N_\delta = 7.48$, $Q = 0.98$ (right column). }
    \label{cylresults}
\end{figure}

Following \citet{Gillis2019}, the quality of the spatial discretization is measured with the two parameters
\begin{equation}
    N_\delta = \frac{1}{\sqrt{\mathrm{Re}_D}}\qty(\frac{D}{h}), \quad\quad Q = \Re_D^{3/2} \qty(\frac{h}{D})^2 = \frac{\Re_D^{1/2}}{N_\delta^2}.
\end{equation}
The parameter $N_\delta$ estimates the number of grid points contained within the characteristic boundary layer thickness, while the parameter $Q$ estimates the mesh Reynold's number based on the boundary vorticity. Gillis' results indicate that $N_\delta \sim 8$ or higher represents a well-resolved flow-field and is generally sufficient for obtaining accurate drag forces, while $Q \sim 1$ or lower allows for accurate wall vorticity values. Figure \ref{cylresults} plots the drag coefficient $C_D = 2F/u_\infty^2 D$ as a function of the non-dimensional time $t^* = u_\infty t/D$ at two Reynolds numbers, $\Re = 550$ and $\Re = 3000$, for spatial resolutions $N_\delta = 8.73$ ($Q = 0.31$) and $N_\delta = 7.48$ ($Q = 0.98$) respectively. The drag coefficients are calculated using a control volume approach, and are in close agreement with results from the immersed interface method of \citet{Gillis2019} and the vortex method of \citet{Koumoutsakos1995}.

Also shown are instantaneous profiles of the non-dimensional surface vorticity $\omega^* = \omega D/U$ as a function of the angular coordinate $\theta$ on the cylinder surface ($\theta = 0$ corresponds to the leading stagnation point.) These distributions are taken from the vorticity boundary condition prescribed during the transport step of the discretization. The present vorticity profiles follow closely the results of \citet{MarichalThesis} at $\Re = 550$ and \citet{Wu2019} at $\Re = 3000$. In both cases the profiles also agree well with \citet{Gillis2019}.

\begin{figure}
    \centering
    \begin{subfigure}{0.46\linewidth}
        \centering
        \resizebox{\textwidth}{!}{\input{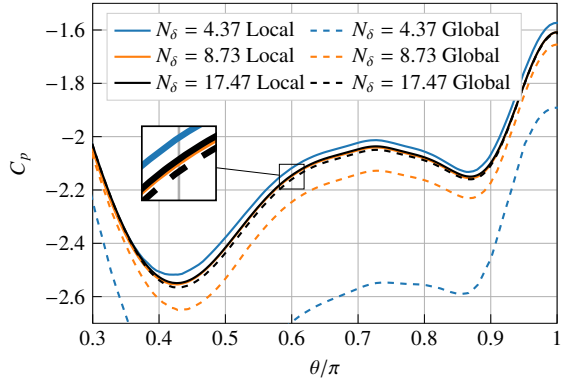}}
        \caption{Resolution effects on pressure calculation for $\Re = 550$ and $t^* = 2.5$. The resolutions considered are $N_\delta = (4.37, \, 8.73, \, 17.47)$ and $Q = (1.23, \, 0.31, \, 0.08)$.}
        \label{cylres550}
    \end{subfigure}
    \hspace{0.03\textwidth}
    \begin{subfigure}{0.46\linewidth}
        \centering
        \resizebox{\textwidth}{!}{\input{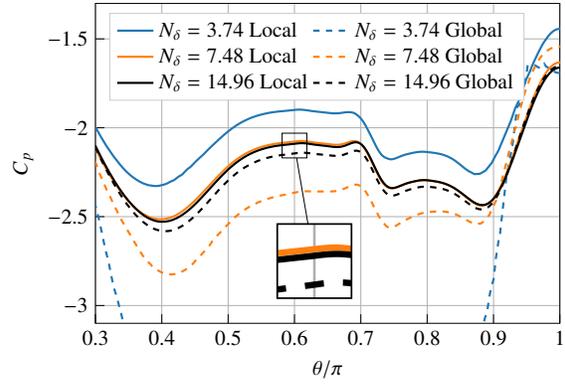}}
        \caption{Resolution effects on pressure calculation for $\Re = 3000$ and $t^* = 2.5$. The resolutions considered are $N_\delta = (3.74, \, 7.48, \, 14.96)$ and $Q = (3.92, \, 0.98, \, 0.24)$.}
        \label{cylres3000}
    \end{subfigure}
    \par\vspace{20pt}   
    \begin{subfigure}{0.46\linewidth}
        \centering
        \resizebox{\textwidth}{!}{\input{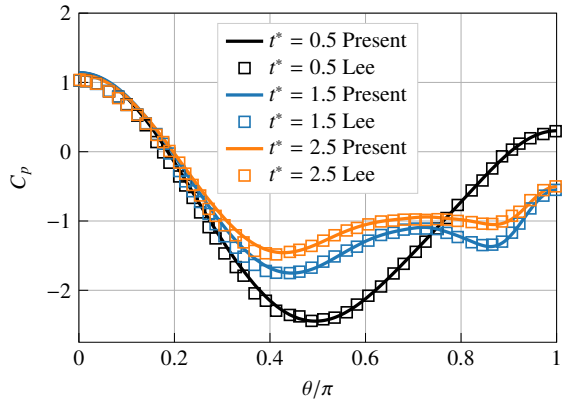}}
        \caption{Surface pressure (global method), $\Re = 550$, $N_\delta = 17.47$.}
        \label{cylpress550}
    \end{subfigure}
    \hspace{0.03\textwidth}
    \begin{subfigure}{0.46\linewidth}
        \centering
        \resizebox{\textwidth}{!}{\input{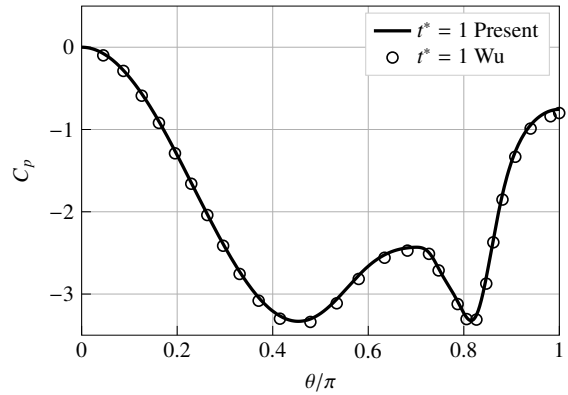}}
        \caption{Surface pressure (local method), $\Re = 3000$, $N_\delta = 14.96$.}
        \label{cylpress3000}
    \end{subfigure}
    \caption{Surface pressure calculations for the impulsively started cylinder at $\Re = 550$ (left column) and $Re = 3000$ (right column). }
\end{figure}

As described in section \ref{section:forces}, the pressure distribution on the cylinder can be calculated either by integrating the surface vorticity flux (a procedure local to each immersed body) or by solving a pressure Poisson equation (a global elliptic solve) . Figures \ref{cylres550} and \ref{cylres3000} show the relative pressure coefficient $C_p(\theta) = 2(p(\theta) - p(0))/u_\infty^2$ resulting from both methods at several spatial resolutions, for $\Re = 500$ and $\Re = 3000$ at $t^* = 2.5$.  At comparable values of the $N_\delta$ parameter, the convergence of both the local and global pressure calculations at $\Re = 3000$ is slower than at $\Re = 550$. The convergence is more consistent across Reynolds numbers when evaluated with the $Q$ parameter: for both Reynolds numbers, $Q \sim 0.3$ is sufficient for a well-converged local pressure calculation. Similar convergence in the global pressure calculation requires even more resolution, and is only achieved in the finest resolution at $\Re = 550$ ($Q = 0.08$). Figures \ref{cylpress550} and \ref{cylpress3000} demonstrate that the converged global pressure at $\Re = 550$ and local pressure at $\Re = 3000$ agree well with reference data from the vorticity-based Brinkmann penalization method of \citet{Lee2014} and the vorticity-based body-fitted finite volume method of \citet{Wu2019}.

\subsection{Semicircular lid-driven cavity}
\begin{figure}
    \centering
    \includegraphics[width=0.5\textwidth]{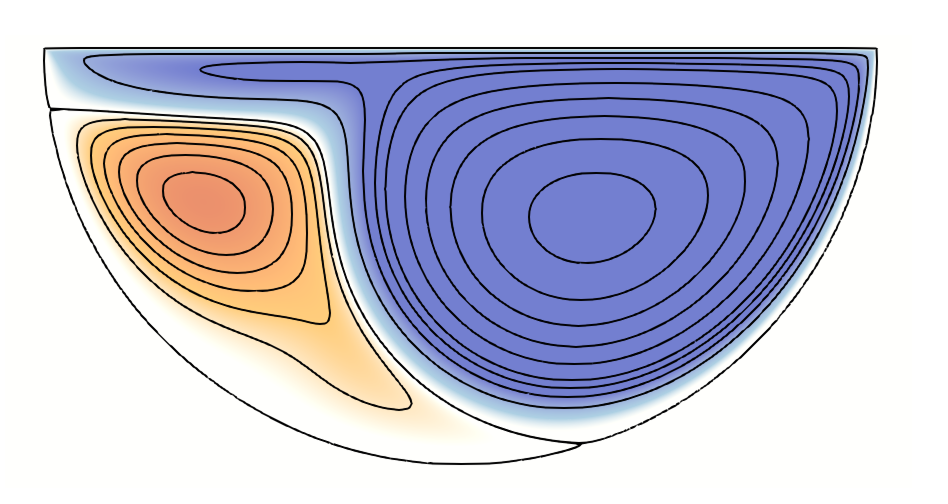}
    \caption{Steady-state streamlines for the flow in a semicircular lid-driven cavity, colored by stream function ($\Re_D = 3000$). Contours indicate $\psi/UD = (-6\times10^{-3}, \ -5\times10^{-3}, \ -4\times10^{-3}, \ -3\times10^{-3}, \ -2\times10^{-3}, \ -1\times10^{-3}, \ -1\times10^{-6}, \ 1\times10^{-6}, \ 5\times10^{-3}, \ 1\times10^{-2}, \ 1.5\times10^{-2}, \ 2\times10^{-2}, \ 3\times10^{-2}, \ 4\times10^{-2}, \ 5\times10^{-2}, \ 6\times10^{-2}, \ 7\times10^{-2})$.}
    \label{fig:semisetup}
\end{figure}

\begin{figure}
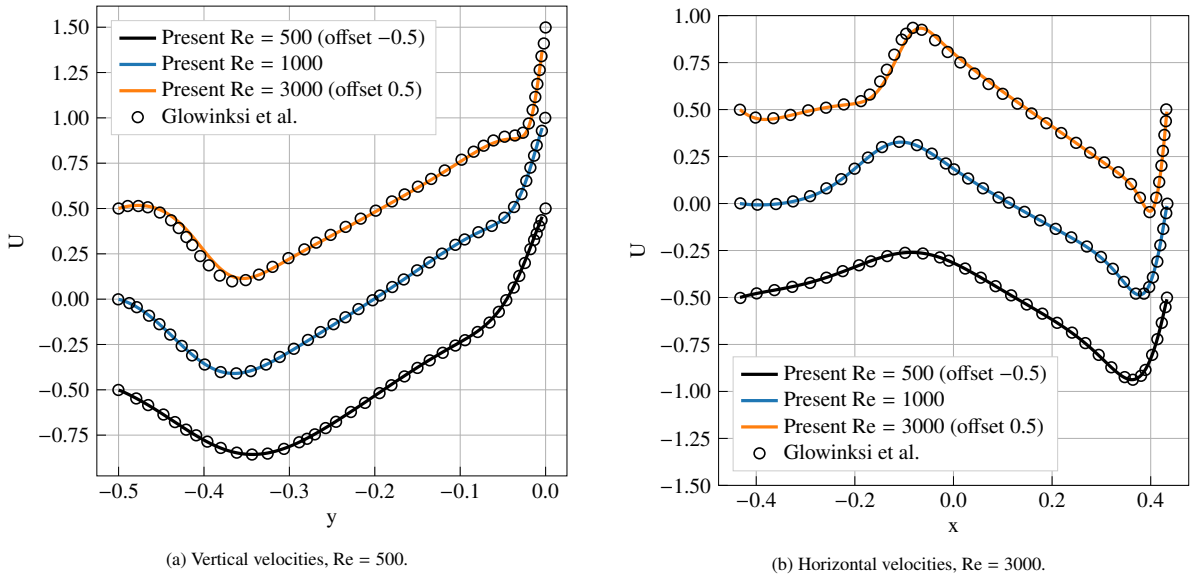

    \centering
    \begin{subfigure}{0.46\textwidth}
        \centering
        \resizebox{\textwidth}{!}{\input{tikzfigures/Semicircular_Vertical}}
        \caption{Vertical velocities, $\Re = 500$.}
        \label{fig:scv}
    \end{subfigure}
    \hspace{0.03\textwidth}
    \begin{subfigure}{0.46\textwidth}
        \centering
        \resizebox{\textwidth}{!}{\input{tikzfigures/Semicircular_Horizontal}}
        \caption{Horizontal velocities, $\Re = 3000$.}
        \label{fig:sch}
    \end{subfigure}
    
    \caption{Steady-state velocity profiles from the interior of a lid-driven semicircular cavity. Data are offset for readability. Both the (a) vertical (b) and horizontal profiles show excellent agreement with reference data from \citet{Glowinski2006} for Reynolds numbers between 500 and 3000.}
    \label{fig:semicavity}
\end{figure}

While the method developed here has several computational advantages for exterior flows, it is equally capable of simulating internal flows with concave boundaries. Consider a semicircular cavity with diameter $D$ filled with a stationary fluid of viscosity $\nu$ (Figure \ref{fig:semisetup}). At $t = 0$ the top wall of the cavity begins moving rightward with velocity $U$. The resulting flow is characterized by the Reynolds number $\Re = UD/\nu$, and has been shown to reach a steady state for $\Re \le 6600$ \citep{Glowinski2006}. To simulate this flow numerically, the semicircular flow domain is embedded in a rectangular computational domain, and the center of the semicircle offset from the grid to break symmetry. The resulting flow is integrated in time until it reaches an approximate steady state. Velocity profiles taken from this steady flow are shown in Figure \ref{fig:semicavity}, and show excellent agreement with those provided by Glowinski et al. \citep{Glowinski2006} for $\Re = 500$, $1000$, and $3000$.
 
For internal flows, the boundary of the computational domain lies outside of the fluid domain. As a result, the convolution operator $\nabla_h^{-2}$ used to solve the velocity reconstruction problem need not satisfy a particular far-field boundary condition. Here we choose the Dirichlet boundary condition $\psi = 0$, which is the most convenient and least computationally expensive option \citep{Caprace2021}.

\subsection{Side-by-side cylinder pairs}
\begin{figure}[htb]
    \centering
    \includegraphics[width=\textwidth]{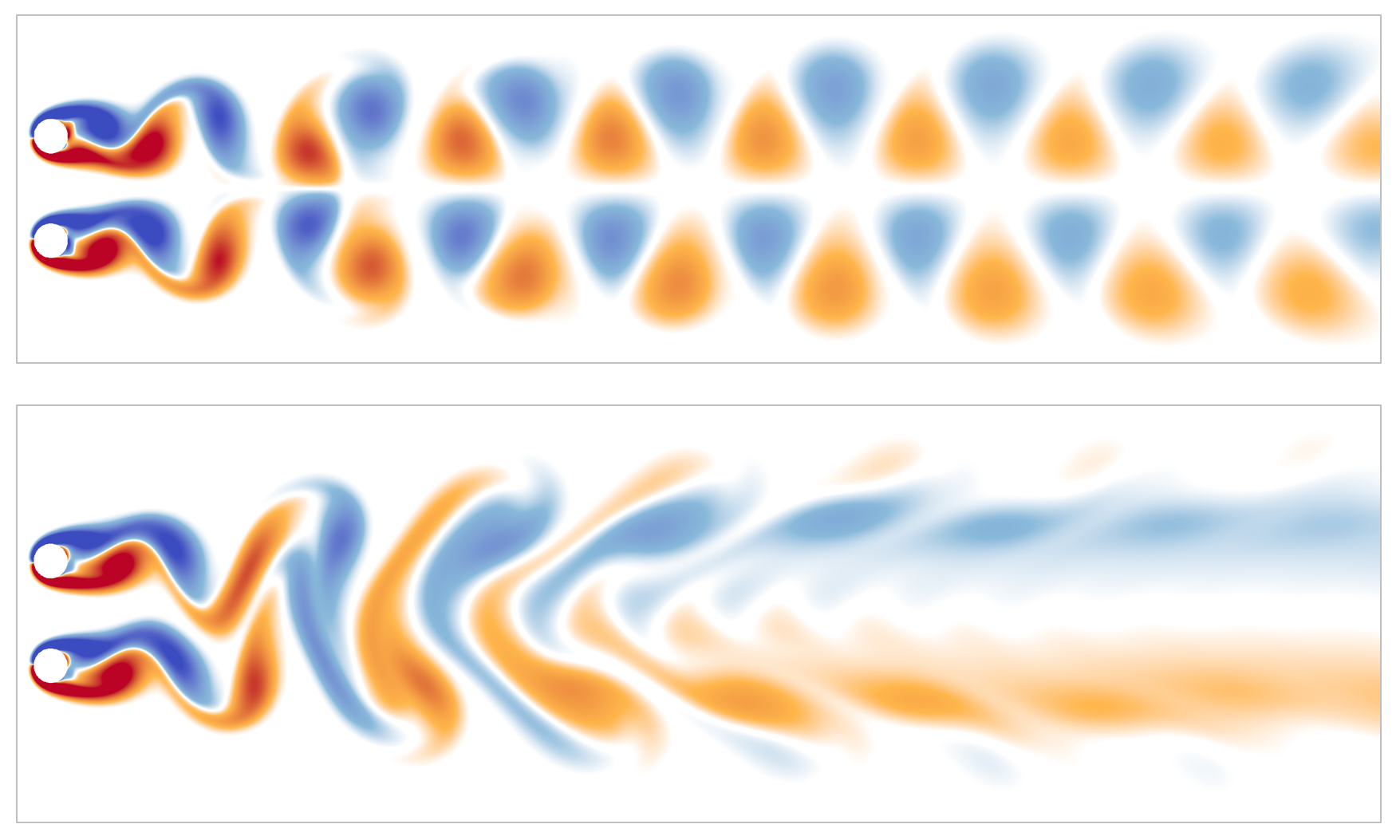}
    \caption{Two distinct shedding patterns for a side-by-side cylinder pair at $\Re = 100$ and $L/D = 3$: antiphase shedding (top) and in-phase shedding (bottom).}
    \label{fig:sidebyside}
\end{figure}

To validate the ability of our method to simulate flow past multiple bodies, we consider a side-by-side cylinder pair. In this test case, two cylinders of diameter $D$ are placed side-by-side in a free stream flow of velocity $U$, with centers separated by a distance $L$. The flow is characterized by two non-dimensional parameters, the Reynolds number $\Re = UD/\nu$ and the non-dimensional gap-width $L/D$. For certain combinations of these two numerical parameters,  multiple stable vortex-shedding modes exist \citep{Kang2003}; here we consider in-phase and antiphase shedding, illustrated in Figure \ref{fig:sidebyside}. Both patterns can be reached from a null initial vorticity field, with antiphase shedding coming from a constant free stream and in-phase shedding coming from a free stream which is initially perturbed to break symmetry.

\begin{figure}[htb]
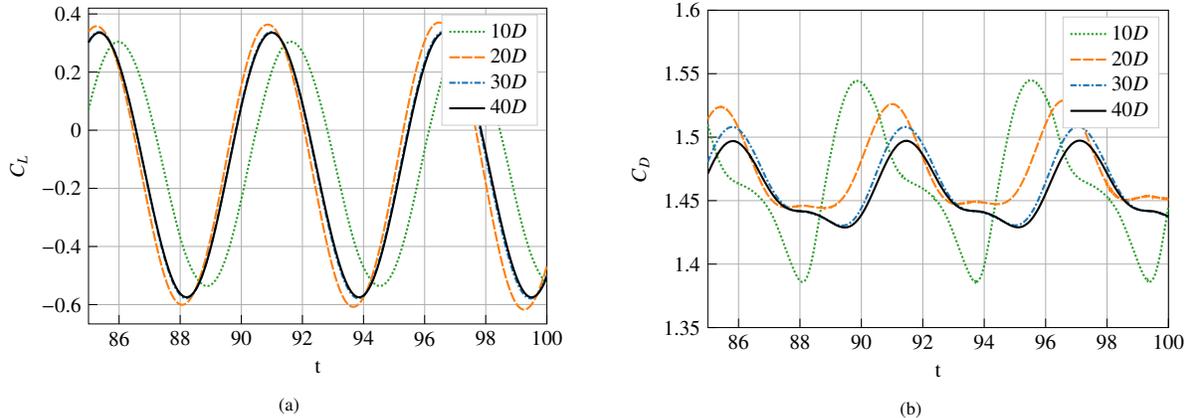

    \centering
    \begin{subfigure}{0.46\textwidth}
        \centering
        \resizebox{\textwidth}{!}{\input{tikzfigures/outflow_lift}}
        \caption{}
    \end{subfigure}
    \hspace{0.03\linewidth}
    \begin{subfigure}{0.46\linewidth}
        \centering
        \resizebox{\textwidth}{!}{\input{tikzfigures/outflow_drag}}
        \caption{}
    \end{subfigure}
    \caption{Time history of the (a) lift coefficient and (b) drag coefficients of the lower cylinder in a cylinder pair at $\Re = 100$ and $L/D = 3$, for different domain lengths. The calculations performed on domains of length $30D$ and $40D$ agree well; shorter domains lead to slight shifts in the amplitude and phase of the lift coefficient, and large qualitative changes in the behavior of the drag coefficient. }
    \label{fig:outflowlocation}
\end{figure}

To simulate this test case, we place the center of each cylinders a distance $1D$ from the inflow boundary. The lack of padding at the inflow is enabled by the true free-space boundary conditions implemented in the velocity reconstruction procedure \citep{Caprace2021}. At $t = 0$ the flow starts from a null vorticity field, and is integrated in time until a steady-state shedding pattern is reached. The outflow boundary condition described in section \ref{section:outflow} is prescribed on the downstream domain boundary to allow for long-time integration. An appropriate location for the outflow boundary can be determined by calculating the lift and drag forces resulting from a single set of parameters ($\Re = 100$, $L/D = 3$) and a varied domain length. The results, shown in Figure \ref{fig:outflowlocation}, indicate that these forces are relatively insensitive to outflow location for domains longer than $30D$, and to allow a margin of safety we adopt a domain of size $12D \times 40D$. For all of the simulations shown here, the spatial resolution has been chosen to ensure accurate pressure calculations via surface integration. Using the resolution parameters defined in section \ref{section:impcylresults}, this corresponds to $Q = 0.17$ at $\Re = 100$ and $Q = 0.34$ at $\Re = 160$.

\begin{figure}[htb]
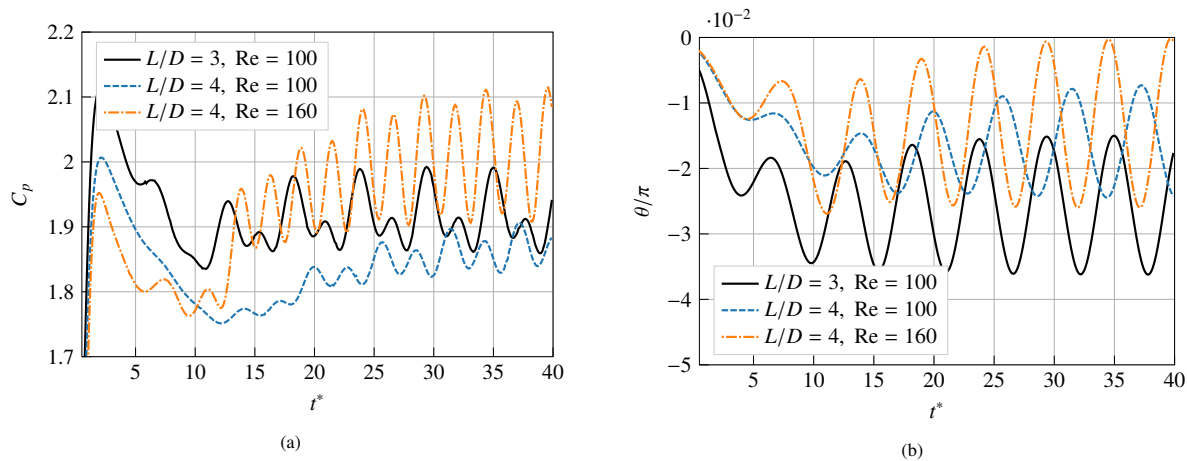

    \begin{subfigure}{0.46\textwidth}
        \centering
        \resizebox{\textwidth}{!}{\input{tikzfigures/cylpair_pressure_trace}}
        \caption{}
        \label{fig:cylpairpress}
    \end{subfigure}
    \hspace{0.03\textwidth}
    \begin{subfigure}{0.46\textwidth}
        \centering
        \resizebox{\textwidth}{!}{\input{tikzfigures/cylpair_stagnation_trace}}
        \caption{}
        \label{fig:cylpairstag}
    \end{subfigure}
    
    \caption{(a) Time history of the pressure difference between the leading stagnation point and rear $\theta = \pi$ point for cylinder pairs undergoing antiphase shedding at various gap widths and Reynolds numbers. (b)  The time history of the location of the leading stagnation point on the lower cylinder for the same set of parameters considered on the left.}
    \label{fig:cylpair_pressure}
\end{figure}

Table \ref{tab:sidebyside} lists the steady-state statistics of the drag coefficient $C_D = 2F_x/DU^2$ and lift coefficient $C_L = 2F_y/DU^2$ of each cylinder for a variety of Reynolds numbers, gap widths, and shedding patterns. Reference values are provided by \citet{Kang2003}, who uses a velocity-pressure immersed boundary method with an $80D \times 100D$ computational domain. The two sets of results show good agreement, despite the fact the present method uses a domain that is sixteen times smaller in area. In addition to lift and drag forces, the use of a sharp immersed method allows for the calculation of time-dependent pressure distributions on the immersed cylinders. Figure \ref{fig:cylpairpress} displays the time history of $C_p = 2(p_{stag} - p(\pi))/u_\infty^2$, the difference between the pressure at the leading stagnation point and the $\theta = \pi$ point on the downstream side of each cylinder, for cylinder pairs undergoing antiphase shedding at three different sets of parameters. For these flows the leading stagnation point is identified as the point of zero vorticity closest to $\theta = 0$ on each cylinder, which is calculated from the time-dependent surface vorticity distribution. The time history of the angular location of this point on the lower cylinder of each pair is provided in Figure \ref{fig:cylpairstag}.

\begin{table}[htb]
    \centering
    \begin{tabular}{ | c | c | c | c | c | } 
    \hline
    Parameters & Author & $C_{D, Mean}$ & $C_{L, Mean}$ & $C_{L, RMS}$ \\ 
    \hline\hline
    \multirow{2}{*}{$\Re = 100$, $L/D = 3$, Antiphase}
    & Kang & 1.46 & 0.116 & 0.317 \\
    & Present & 1.47 & 0.129 & 0.319 \\ 
    \hline
    \multirow{2}{*}{$\Re = 100$, $L/D = 3$, In-Phase}
    & Kang & 1.44 & 0.129 & 0.190 \\
    & Present & 1.42 & 0.120 & 0.183 \\ 
    \hline
    \multirow{2}{*}{$\Re = 100$, $L/D = 4$, Antiphase}
    & Kang & 1.43 & 0.082 & 0.280 \\
    & Present & 1.42 & 0.070 & 0.273 \\
    \hline
    \multirow{2}{*}{$\Re = 160$, $L/D = 3$, Antiphase}
    & Kang & 1.45 & 0.100 & 0.510 \\
    & Present & 1.44 & 0.092 & 0.507 \\ 
    \hline
    \multirow{2}{*}{$\Re = 160$, $L/D = 4$, Antiphase}
    & Kang & 1.40 & 0.058 & 0.440 \\
    & Present & 1.39 & 0.056 & 0.443 \\ 
    \hline
    \end{tabular}
    \caption{Long-time statistics of the lift and drag forces on a side-by-side cylinder pair, as calculated \citet{Kang2003} and by the present method.}
    \label{tab:sidebyside}
\end{table}

\subsection{Multiple non-convex immersed bodies}
\begin{figure}
    \centering
    \includegraphics[width=\textwidth]{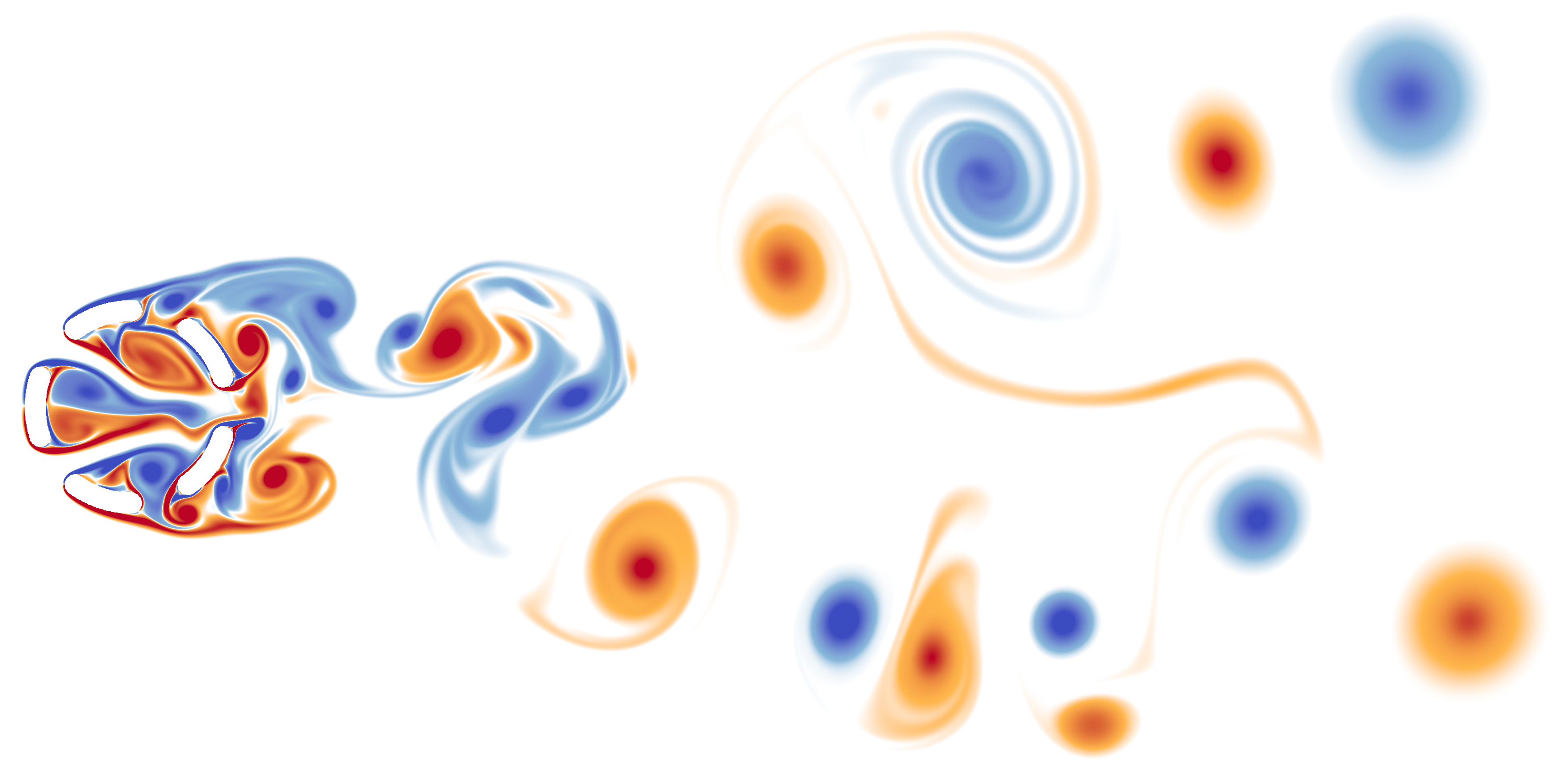}
    \caption{A snapshot of the vorticity field resulting from flow over a group of non-convex solid bodies inspired by a submerged offshore aquaculture cage structure.}
    \label{fig:showcase}
\end{figure}

We finally demonstrate the flexibility of this framework through a test case that combines multiple non-convex obstacles in an external flow. Figure \ref{fig:showcase} provides a snapshot of the vorticity field that results from an impulsively started flow over a collection of solid bodies inspired by a submerged offshore aquaculture cage structure. The circulation around each body is automatically tracked by our solver, and our IIM is robust to the geometric issues caused by concave surfaces (as discussed in section \ref{section:2diim}).

\section{Conclusion}\label{section:conclusion}
We have presented a 2D vorticity-based immersed interface method that can simulate fluid flows in bounded and unbounded domains, with multiple non-convex immersed bodies, and outflow boundary conditions. Our approach relies on a re-interpretation of the explicit jump IIM which simplifies the implementation and addresses challenges posed by nonconvex bodies. We show that the use of conservative spatial discretizations allows for the discrete enforcement of Kelvin's theorem, which is essential for simulations with multiple bodies and outflow boundary conditions. Lastly, we have built upon an efficient FFT-accelerated elliptic solver to solve the velocity reconstruction problem with multiple immersed bodies on arbitrary domain topologies. On test cases with known solutions, the resulting method achieves second order spatial convergence in the infinite error norm over the entire domain, and third order temporal convergence. We reproduce reference results for a variety of internal and external flows with Reynolds numbers between 100 and 3000, and accurately recover lift forces, drag forces, moments, and time-dependent traction distributions on immersed solid bodies. Lastly, we have demonstrated that with free-space and outflow boundary conditions, our vorticity-based approach can recover accurate solutions with a domain size that is sixteen times smaller than that of the velocity-based reference results.

We consider several immediate future directions for this work. Immersed interface methods are well-suited for simulations involving moving and deforming geometry, including fluid-structure interaction problems that are often discretized with lower-order immersed boundary or penalization methods. There are also promising developments in extending the IIM to non-smooth geometries with thin features, cusps, and acute interior corners \citep{Marichal2014, Hosseinverdi2020}, which would further broaden the range of flows that can be simulated with the current method. The accuracy of surface pressure and shear distributions can be greatly increased through the use of multi-resolution adaptive grids \citep{Rossinelli2015}, which allow computational elements to be concentrated around immersed surfaces. Finally, such multiresolution adaptive grids would pave the way for a computationally-efficient extension to 3D, building upon \citep{GillisThesis} as well as the various improvements laid out in this work. 

\section*{Acknowledgements}
We wish to acknowledge financial support from a MathWorks Engineering Fellowship (JG); from a postdoctoral fellowship from the Belgian American Educational Foundation (BAEF) and an International Excellence Scholarship from Wallonie Bruxelles International (TG); from a MIT International Science and Technology Initiatives (MISTI) Seed Fund award (PC,WVR); and from an Early Career Award from the Department of Energy, Program Manager Dr.~Steven~Lee, award number DE-SC0020998 (JG, TG,  WVR).

\bibliographystyle{model1-num-names}
\bibliography{refs}

\appendix

\section{Geometry Processing}\label{section:geometryprocessing}
In the immersed interface method, a solid body is represented entirely by the intersections between a boundary curve and the lines of a Cartesian grid, as well as a normal vector to the boundary at these intersections. Below we present an efficient algorithm which determines these intersections with $\mathcal{O}(h^4)$ accuracy and normal vectors with $\mathcal{O}(h^3)$ accuracy for any object that can be described by a smooth level set. 

Let $\phi(\vb{x})$ be a smooth level set satisfying $\phi(\vb{x}) = 0$ on an immersed boundary, and let $\phi_{ij}$ be its value at the grid point $\vb{x}_{ij}$. Each control point corresponds to a pair of neighboring points $\vb{x}_{ij}$ and $\vb{x}_{kl}$ for which $\phi_{ij} < 0$ and $\phi_{kl} \ge 0$: because $\phi(\vb{x})$ is continuous, there is a control point $\vb{x}_c$ on the grid-line connecting $\vb{x}_{ij}$ and $\vb{x}_{kl}$ for which $\phi(\vb{x}_c) = 0$. To locate this intersection efficiently, we limit our attention to the one-dimensional function $\tilde{\phi}(z) = \phi(\vb{x}_{ij} + z(\vb{x}_{kl} - \vb{x}_{ij}))$, which restricts $\phi(\vb{x})$ to the grid line connecting $\vb{x}_{ij}$ and $\vb{x}_{kl}$. 

\begin{figure}
    \centering
    \includegraphics[width=0.45\textwidth]{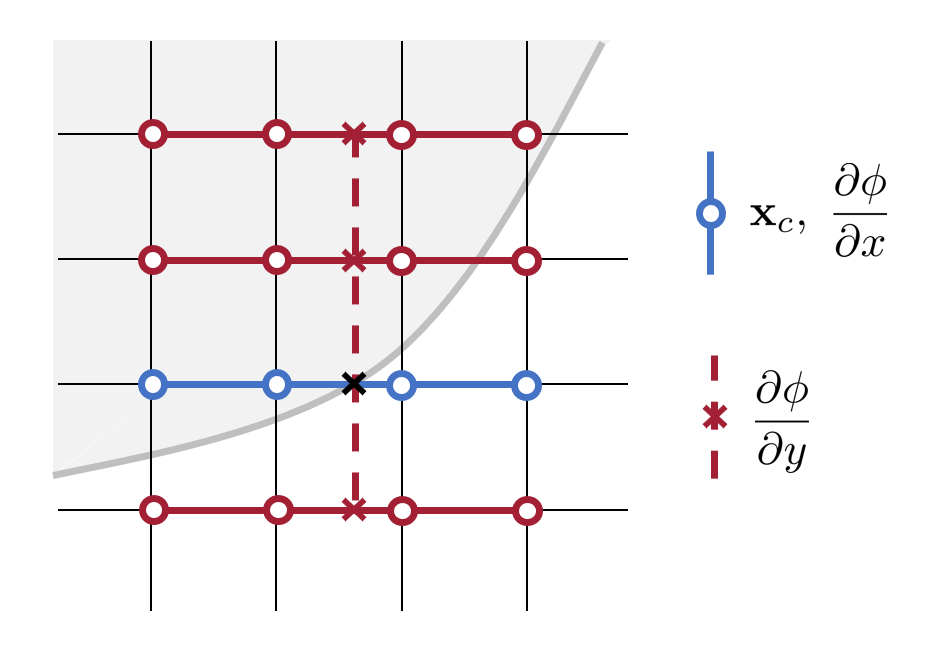}
    \caption{Stencils used to calculate intersection points and normal vectors. The intersection (black cross) is located by fitting a cubic polynomial to the four level set values shown in blue. To calculate the gradient of the level set, the x-direction derivative is taken with the stencil shown in blue, while the y-direction derivative is taken with the stencil shown in red crosses. Each red cross level set value is interpolated using a four point stencil shown in red circles. }
    \label{fig:normalvectors}
\end{figure}

To avoid any additional evaluations of $\phi$, we will find the roots of a polynomial which interpolates $\tilde{\phi}(z)$. As a first approximation, we find the root of a linear interpolating polynomial, giving $z_0 = -\phi_{kl}h / (\phi_{kl} - \phi_{ij})$. This approach locates $\vb{x}_c$ with second order accuracy. To improve on this, we construct a cubic polynomial $p_3(z)$ which interpolates $\tilde{\phi}(z)$ at $z = mh$ for integers m satisfying $-1 \le m \le 2$. One root ($z_1$) can be found using Newton's method with $z_0$ as an initial guess, and it is likely to lie in $[0, h]$. If it does not, we extract a quadratic factor from $p_3$, so that 
\begin{align*}
    p_3(z) &= a_3z^3 + a_2z^2 + a_1z + a_0 \\
           &= (z-z_1)(b_2z^2 + b_1z + b_0), \qq{with} \\
    &\begin{cases} b_2 &= a_3, \\
    b_1 &= a_2 + z_1b_2, \\
    b_0 &= a_1 + z_1b_1. \end{cases} 
\end{align*}
The remaining two roots of $p_3$ can then be found with the quadratic formula. One of these roots lies in $[-h,0]$, and provides a fourth order estimate of the location of $\vb{x}_c$. The normal vector $\vu{n} = \nabla \phi(\vb{x}_c)/|\nabla \phi(\vb{x}_c)|$ can then be calculated with third order accuracy using four-point finite difference stencils and interpolations, as shown in Figure \ref{fig:normalvectors}.

\section{Stability of the Transport Discretization}\label{section:transportstability}
\begin{figure}[htb]
    \centering
    \begin{subfigure}{0.49\linewidth}
        \centering
        \includegraphics[width=\textwidth]{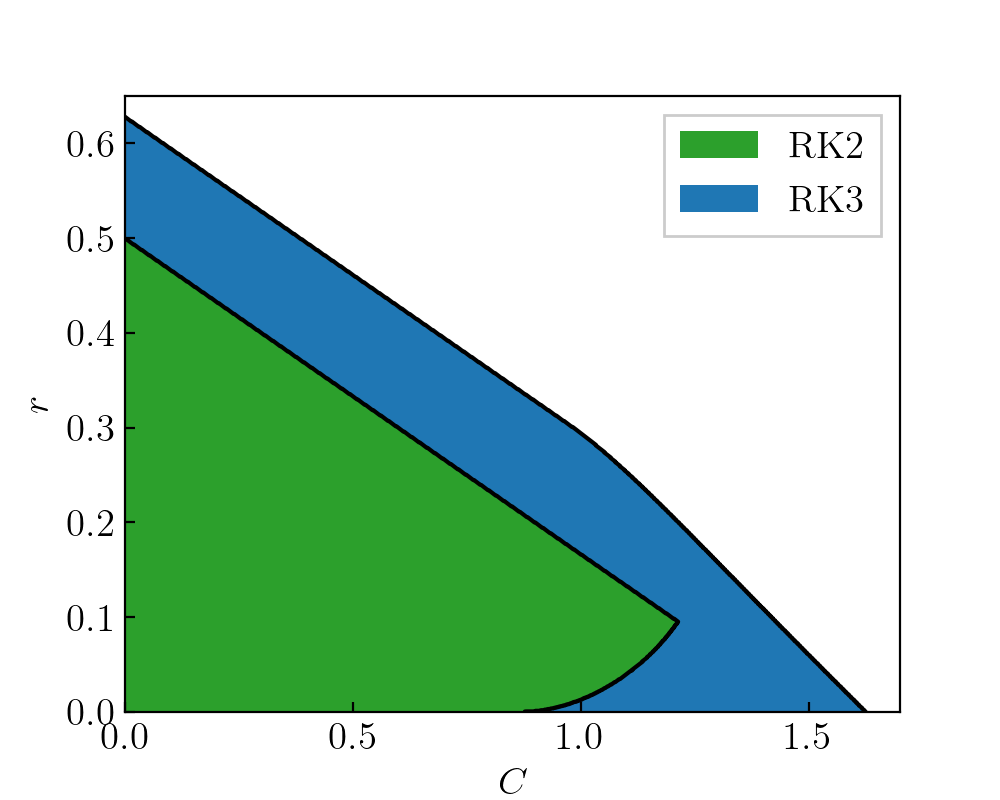}
        \caption{1D Stability Regions.}
        \label{fig:oneDstab}
    \end{subfigure}
    \begin{subfigure}{0.49\linewidth}
        \centering
        \includegraphics[width=\textwidth]{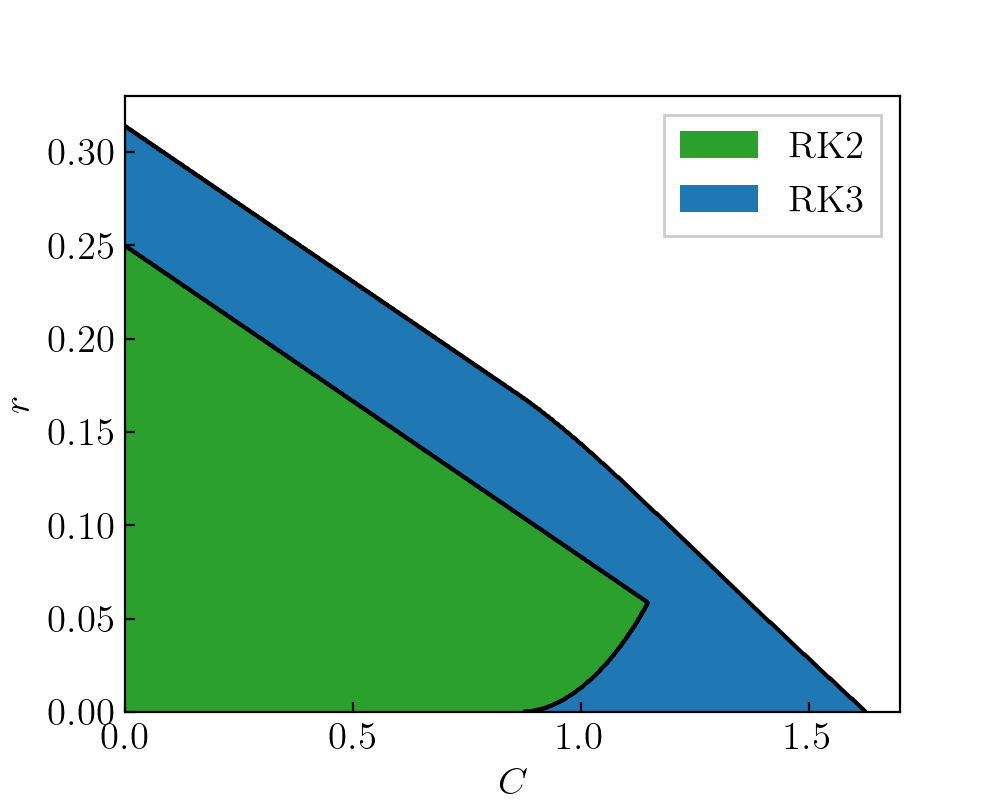}
        \caption{2D Stability Regions.}
        \label{fig:twoDstab}
    \end{subfigure}
    \caption{Stability regions for the 1D and 2D transport discretizations with second and third order Runge-Kutta time integration.}
    \label{fig:transportStab}
\end{figure}

The stability of the free-space transport schemes developed in section \ref{section:transport} can be characterized with a von-Neumann stability analysis, where the velocity field $\vb{u}$ as well as the viscosity $\nu$ are assumed to be constant in space. For a discretization with grid spacing $h$ and time step $\tau$, stability is dependent on the Fourier number $r = \nu \tau/h^2$  and the Courant number $C_{1D} = \lvert u \rvert \tau/h$ for 1D simulations or $C_{2D} = (\lvert u_x \rvert + \lvert u_y \rvert)\tau/h$ for 2D simulations. Stability regions in the $(C, r)$ plane for both the 1D and 2D transport schemes with a third order Runge-Kutta time integration scheme are shown in Figure \ref{fig:transportStab}. For completeness we also provide the stability regions for a second order Runge-Kutta scheme, although we do not use it in this work. 

Neither the 1D or 2D stability region can be expressed as uncoupled constraints $0 \le C \le C_{max}$ and $0 \le r \le r_{max}$. Consequently, its necessary to consider both parameters simultaneously when determining the maximum allowable time step for a given grid resolution. This can be done by noting that the ratio $C/r = Re_h$ is independent of the time step $\Delta t$, so that varying $\Delta t$ traces out a straight line through the origin of the $(C,r)$ plane. The maximum allowable time step $\Delta t_{\mathrm{max}}$ corresponds to the intersection between this line and the boundary of the stability region, which can be calculated easily if the boundary is approximated by a series of linear segments. The vertices of a polygon which reasonably approximates each 2D stability region are listed in Table \ref{tab:stability}. 

All of the time-dependent calculations presented in this work use a fixed fraction of the maximum stable time step $\Delta t = C_{stab} \Delta t_{\mathrm{max}}$, and the constant $C_{\mathrm{stab}}$ is referred to as a ``safety factor" in the main text. For the special case of impulsively started flows, $C_{\mathrm{stab}}$ is greatly reduced at the start of each simulation to resolve the initial dynamics, then smoothly brought back to its prescribed value.

\begin{table}[h]
    \centering
    \begin{tabular}{ | c | c | } 
    \hline
    Time Integration & Vertices $(C,r)$ \\ 
    \hline\hline
    RK3 & (0,0), (1.620, 0),  (0, 0.314) \\
    \hline
    RK2 & (0,0), (0.874, 0), (1.140, 0.058), (0, 0.250) \\
    \hline
    \end{tabular}
    \caption{Stability region of the 2D transport discretization in the $(C,r)$ plane, approximated by simple polygons.}
    \label{tab:stability}
\end{table}

\section{Force Calculation}\label{appendix:forces}

In a vorticity-based solver, calculating forces and tractions on immersed bodies is complicated by the fact that the pressure field is not immediately available. This section brings together a collection of useful results from across the literature which allow 2D vorticity-based methods to circumvent this issue \cite{Noca1997, Zhang2009, Lee2014}. This includes results which are likely known but difficult to find in print, like \eqref{eq:momentidentity} and \eqref{eq:pressuremoment}, as well as a control volume formulation for calculating moments from only the velocity and vorticity fields \eqref{Mm1}.

\subsection{Global Forces}
For a stationary body immersed in a two-dimensional flow, the surface traction vector $\vb{t}$ is related to the pressure and the vorticity through
\begin{equation}
    \vb{t} = -p \vu{n} + \nu \omega \vu{s},
\end{equation}
where $\vu{s} = \vu{k} \times \vu{n}$ is the tangential unit vector. Integrating the traction $\vb{t}$ over the surface of a solid body yields the total force vector $\vb{F}$, while integrating $(\vb{x} - \bar{\vb{x}}) \times \vb{t}$ yields the total moment $\vb{M}_{\bar{\vb{x}}}$ taken about the point $\bar{\vb{x}}$. These integrals can be broken into two contributions: one from the tangential viscous component $\vb{t}_v = \nu \omega \vu{s}$, and one from the normal pressure component $\vb{t}_p = -p\vu{n}$. The viscous contribution can be calculated directly from the surface vorticity field, giving 
\begin{align}
    \vb{F}_v &= \nu \oint_S \omega \vu{s} \dd{s}, \\
    \vb{M}_{v,\vb{\bar{x}}} &= \nu \oint_S \vb{(x - \bar{x})} \times \omega \vu{s} \dd{s}.
\end{align}
The pressure contribution can be obtained from the Navier-Stokes equations applied to a no-slip boundary, 
\begin{equation}\label{pressuregrad}
     0 = - \nabla p - \nu \nabla \times \omega,
\end{equation}
which relates the surface pressure gradient directly to the surface vorticity gradient. Using the integral identities
\begin{equation}
    \oint_S p\vu{n} \dd{s} = -\oint_S \vb{x} \times (\vu{n} \times \nabla p) \dd{s}, \qq{and}
    \oint_S \vb{x} \times p\vu{n} \dd{s} = \oint_S \frac{\abs{\vb{x}}^2}{2} \vu{n} \times \nabla p \dd{s},\label{eq:momentidentity}
\end{equation}
the total pressure loads on the body can be written in terms of the normal vorticity gradient: 
\begin{align}
    \vb{F}_p &= \nu \oint_S (\vb{x} \times \vu{k}) \pdv{\omega}{n} \dd{s}, \\
    \vb{M}_{p,\vb{\bar{x}}} &= \frac{\nu}{2} \oint_S \abs{\vb{x - \bar{x}}}^2\vu{k}\pdv{\omega}{n} \dd{s}.\label{eq:pressuremoment}
\end{align}
Generally, calculations performed on an immersed surface can be significantly noisier than calculations performed on the regular grid. This is compounded by the existence of thin boundary layers in high Reynolds number flows, which lead to large gradients in both velocity and vorticity near no-slip boundaries. Consequently, it is advantageous to have an alternative strategy for calculating the total force and moment on an immersed body that avoids the use of surface quantities. In a vorticity-velocity formulation, this can be done using a control volume approach developed by Noca \cite{Noca1997}, which does not require explicit knowledge of the the pressure field. Noca's ``momentum 4" formulation, specialized to a stationary immersed obstacle with no-slip boundaries and a stationary 2D control volume, is
\begin{equation}\label{Fm}
    \vb{F} = -\dv{}{t}\int_V \vb{u} \dd{V}  - \dv{}{t} \oint_{S} \vu{x} \times (\vu{n} \times \vb{u}) \dd{S} + \oint_S \vu{n} \cdot \bm{\gamma} \dd{S}.
\end{equation}
The quantity $\bm{\gamma}$ is a tensor collecting miscellaneous terms evaluated on the stationary exterior surface of the control volume,
\begin{equation}
    \bm{\gamma} = \frac 12 \abs{\vb{u}}^2\vb{I} - \vb{uu} - \vb{u}(\vb{x} \times \omega \vu{k}) + \vb{x} \cdot (\nabla \cdot \vb{T})\vb{I} - \vb{x}(\nabla \cdot \vb{T}) + \vb{T},
\end{equation}
where $\vb{T} = \nu(\nabla \vb{u} + \nabla \vb{u}^T)$ is the viscous stress tensor. Noca's work does not include an analogous calculation of the moment acting on an immersed body, but it can be derived using similar methods (see \ref{section:nocamoments}). Specializing again to an immersed body with no-slip boundaries and a stationary 2D control volume, the analogous expression is
\begin{equation}\label{Mm1}
    \vb{M} = -\dv{}{t} \int_{V} \vb{x} \times \vb{u} \dd{V}  + \dv{}{t} \oint_{S} \frac{\abs{\vb{x}}^2}{2} \vu{n} \times  \vb{u} \dd{S} + \oint_{S} \bm{\lambda}(\vu{n}) \dd{S}. 
\end{equation}
The quantity $\lambda(\vu{n})$ collects miscellaneous surface terms, and can be written as the action of a tensor $\lambda = \vb{\Lambda} \vu{n}$ if necessary (though it is not convenient to do so here):
\begin{equation}\label{Mm2}
    \bm{\lambda}(\vu{n}) =  \frac{1}{2}\abs{\vb{u}}^2(\vb{x} \times \vu{n}) - (\vb{x} \times \vb{u}) (\vb{u} \cdot\vu{n}) - \frac{\abs{\vb{x}}^2}{2} \vu{n}\times(\vb{u}\times\omega \vu{k}) + \frac{\abs{\vb{x}}^2}{2} (\nabla \cdot \vb{T}) \times \vu{n} + \vb{x}\times \qty(\vb{T}\cdot \vu{n}).
\end{equation}
These expressions provide the total moment about $\bar{\vb{x}} = 0$, and can be shifted to any other center by replacing all occurrences of $\vb{x}$ with $\vb{x} - \bar{\vb{x}}$. The surface integrals in this control volume formulation are taken over a rectangular, axis-aligned box that contains the immersed body, and are discretized with the trapezoidal rule. The volume integrals are taken over the regions inside the box, excluding the inside of the immersed obstacle. These are discretized by combining a polynomial extrapolation with the second order level-set integration method developed by \citet{Towers2009}.

\section{Analytical flow field for the impulsively rotated cylinder}\label{section:improt}
The impulsively rotated cylinder with imposed axisymmetry is a flow problem simple enough to have an analytical solution. An expression for the resulting velocity field is provided by Lagerstrom in \cite{Lagerstrom1996}, but little detail on the solution method is provided, and the accompanying expression for the vorticity field is erroneous. Further, the velocity is given in a form which is difficult to evaluate numerically. Here we re-derive Lagerstrom's velocity field and the correct expression for the corresponding vorticity field, then transform both into a form that lends itself to accurate numerical evaluation. 

\begin{figure}
    \centering
    \includegraphics[width=0.4\textwidth]{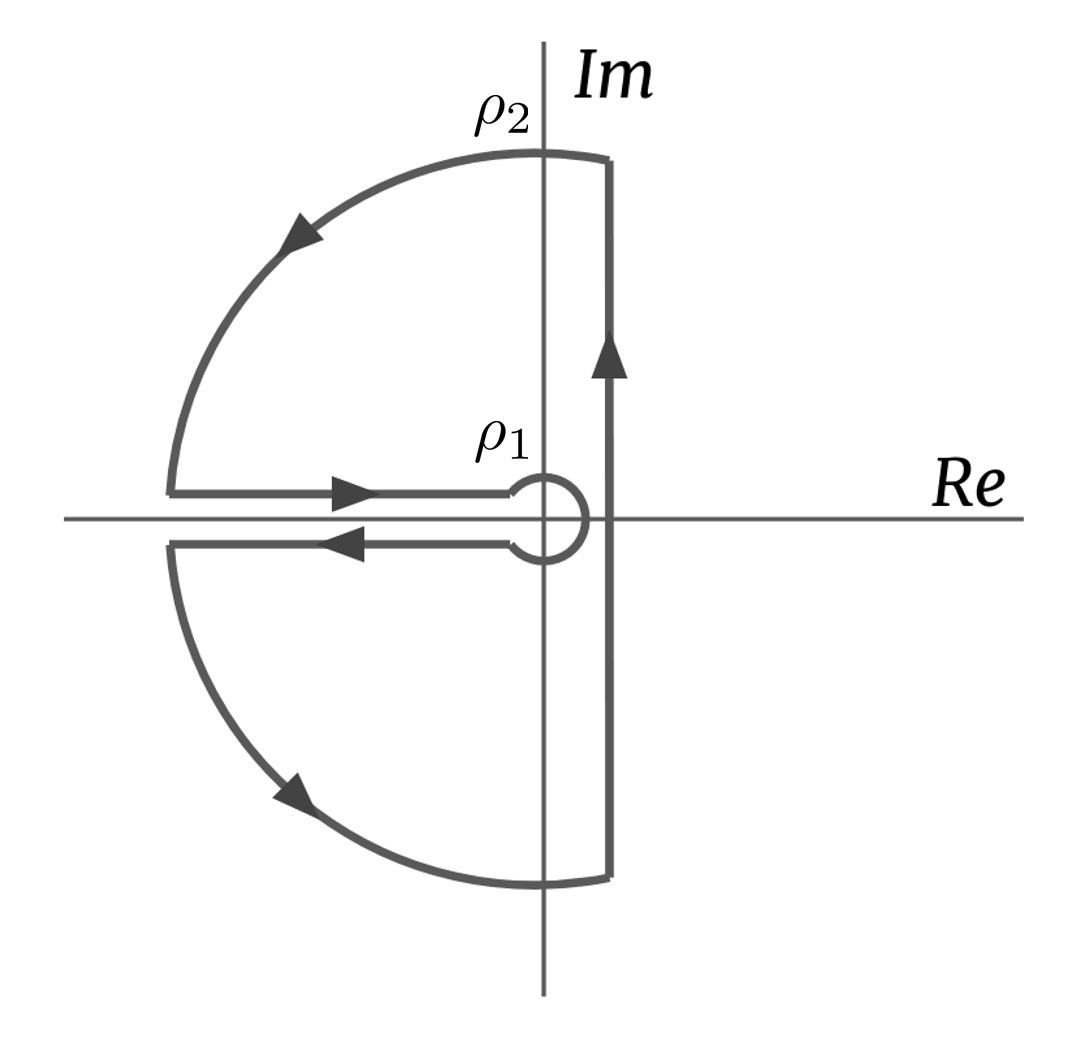}
    \caption{Keyhole contour for the integrand $W(r,s)e^{st}$ in the complex plane.}
    \label{fig:contour}
\end{figure}

Consider a cylinder of radius $R$ at rest in an unbounded fluid domain with viscosity $\nu$. At $t = 0$, the cylinder begins to rotate with angular velocity $\Omega$. For convenience we will work in polar coordinates and define the non-dimensional variables $t^* = \nu t / R^2$, $u^* = u_\theta/\Omega R$,  $r^* = r/R$, and $\omega^* = \dv{u^*}{r^*} = \omega/\Omega$. In these non-dimensional variables, the Navier-Stokes equations with imposed axisymmetry and a no-slip boundary condition reduce to the one-dimensional linear PDE 
\begin{equation}
\begin{gathered}
\pdv{u}{t} =  \pdv[2]{u}{r} + \frac{1}{r}\pdv{u}{r} - \frac{1}{r^2}u, \\[0.5ex]
u(1,t) = 1, \qq{and} \lim_{r\rightarrow\infty}u(r,t) = 0 \\[0.5ex]
u(r,0) = 0,
\end{gathered}
\end{equation}
where we have dropped the asterisks for readability. This problem can be solved with a Laplace transform for the time variable, which leads to an ODE governing the transformed velocity $U(r,s)$:
\begin{equation}
\begin{gathered}
r^2\pdv[2]{U}{r} + r\pdv{U}{r} - (1 + sr^2)U = 0, \\[0.5ex]
U(1,s) = \frac{1}{s}, \qq{} \lim_{r\rightarrow\infty}U(r,s)=0.
\end{gathered}
\end{equation}
The substitution $\beta = \sqrt{s}r$ transforms this into the modified Bessel equation, 
which is solved by a linear combination of the two modified Bessel functions, $U(r,s) = c_1I_1(\beta) + c_2K_1(\beta)$. The arbitrary constants are fixed by applying the boundary conditions, yielding
\begin{equation}
    U(r,s) = \frac{1}{s}\frac{K_1(\sqrt{s}r)}{K_1(\sqrt{s})}.
\end{equation}
This is the expression provided by Lagerstrom. Another quantity of interest is the Laplace transform of vorticity field $W(r,s)$. Using the definition of vorticity in polar coordinates,
\begin{equation}
    W(r,s) = \pdv{U}{r} - \frac{U}{r} = -\frac{1}{\sqrt{s}}\frac{K_0(\sqrt{s}r)}{K_1(\sqrt{s})}.
\end{equation}
Applying an inverse Laplace transform, the time-dependent vorticity field is given by
\begin{equation}\label{eq:improt1}
    \omega(r,t) = - \frac{1}{2\pi i}\int_{\gamma - i\infty}^{\gamma + i\infty} \frac{K_0(\sqrt{s}r)}{K_1(\sqrt{s})} \frac{e^{st}}{\sqrt{s}} \dd{s},
\end{equation}
where $\gamma$ is an arbitrary positive constant. This integrand oscillates with period $2\pi/t$ and decays slowly as $\abs{s} \rightarrow \infty$, making the integral difficult to approximate by conventional methods. To avoid this we choose to integrate over a different contour in the complex plane. The integrand $W(r,s)e^{st}$ has a branch cut on the negative real axis (due to the presence of $\sqrt{s}$ in the expression) and a singularity at $s = 0$, but otherwise is analytic. Consequently, the integral of $W(r,s)e^{st}$ over the closed contour shown in Figure \ref{fig:contour} vanishes. We note that the integral along the arcs in the left half-plane vanishes as $\rho_2 \rightarrow \infty$, while the integral along the contour encircling the origin vanishes as $\rho_1 \rightarrow 0$. Consequently, as the contour grows, the entire integral in \eqref{eq:improt1} is equal and opposite the integrals taken above and below the branch cut: 
\begin{equation}
    \omega(r,t) = \frac{1}{2\pi i} \int_{-\infty+i\epsilon}^{0 + i\epsilon} \frac{K_0(\sqrt{s}r)}{K_1(\sqrt{s})} \frac{e^{st}}{\sqrt{s}}\dd{s} + \frac{1}{2\pi i}\int_{0-i\epsilon}^{-\infty - i\epsilon} \frac{K_0(\sqrt{s}r)}{K_1(\sqrt{s})}\frac{e^{st}}{\sqrt{s}} \dd{s}
\end{equation}
To simplify the integrand further, we make the substitution $\sqrt{s} = -ix$, which ``unfolds" the path of integration to lie along the real axis. The resulting expression is
\begin{equation}
    \omega(r,t) = \frac{1}{\pi}\int_{-\infty} ^{\infty}\frac{K_0(ixr)}{K_1(ix)} e^{-x^2t} \dd{x}.
\end{equation}
This integrand has an even real part and an odd imaginary part, allowing for the simplification
\begin{equation}
    \omega(r,t) = \frac{2}{\pi}\int_0^\infty\real\qty{\frac{K_0(ixr)}{K_1(ix)}}e^{-x^2t}\dd{x}.
\end{equation}
To avoid complex arithmetic, $K_\alpha(ix)$ can be re-expressed as a combination of $J_\alpha(x)$ and $Y_\alpha(x)$, the Bessel functions of the first and second kind, giving
\begin{equation}
    \omega(r, t) = \frac{2}{\pi}\int_0^\infty\frac{J_0(xr)Y_1(x) - J_1(x)Y_0(xr)}{J_1(x)^2 + Y_1(x)^2}e^{-x^2t}\dd{x}.
\end{equation}
This integrand is non-oscillatory, non-singular at $x = 0$, and decays as $e^{-x^2t}$ when $x \rightarrow \infty$. Consequently, it can be evaluated numerically to any desired degree of accuracy.

The velocity field can be evaluated using the same contour described above, with small alterations. The velocity integrand $U(r,s)e^{st}$ has a branch cut on the negative real axis and a singularity at $s = 0$. However, in this case the singularity is proportional to $s^{-1}$, so that the portion of the contour encircling the origin cannot be ignored. Evaluating the additional contribution via residues gives the velocity field 
\begin{align}
    u(r,t) &= \frac{1}{r} + \frac{2}{\pi}\int_0^\infty\imaginary \qty{\frac{K_1(ixr)}{K_1(ix)}}\frac{e^{-x^2t}}{x}\dd{x} \\[0.5ex]
    &= \frac{1}{r} + \frac{2}{\pi}\int_0^\infty \frac{J_1(xr)Y_1(x) - J_1(x)Y_1(xr)}{J_1(x)^2 + Y_1(x)^2}\frac{e^{-x^2t}}{x}\dd{x}.
\end{align}  
Finally, we evaluate the shear stress and total moment acting on the cylinder's surface. Working now with dimensional variables, the shear stress acting at each point on the cylinder's surface has magnitude $\tau(t) = \nu \omega(R,t) - 2\nu\Omega$. The total moment acting on the cylinder is related to the shear stress through $M(t) = 2\pi R^2 \tau(t)$. Defining a non-dimensional moment $M^* = M/2\pi R^2\nu\Omega$ then leads to the straightforward non-dimensional relation
\begin{equation}
    M^*(t^*) = \omega^*(1, t^*) - 2.
\end{equation}

\section{Control volume moment calculation for 3D incompressible flows}\label{section:nocamoments}
There are a variety of methods for using a control volume analysis to obtain the moment acting on an immersed body \cite{Bergmann2011, Nangia2017}.  It is more difficult, however, to find a formulation that makes no assumptions on the size or position of the control volume, and that does not require the pressure field. In \cite{Noca1997}, Noca derives a control volume method satisfying these constraints that determines the force on an immersed body. A similar derivation for moments can be done in a completely analogous way; however, it appears to be absent from the literature. 

Let $V(t)$ be a 3D region which contains on immersed solid body. The internal boundary of $V(t)$ which borders the solid will be denoted $S_b(t)$, with a normal vector that points into $V(t)$ and out of the solid, while the external boundary will be denoted $S(t)$ with a normal vector that points out of $V(t)$. We begin with the conservation of angular momentum for V(t),
\begin{multline}\label{eq:angmom}
    \vb{M}  = -\dv{}{t}\int_{V(t)} \vb{x} \times \vb{u} \dd{V} + \oint_{S(t)} \vb{x} \times [(-p\vb{I} + \vb{T})\cdot \vu{n}] \dd{S} \\  - \oint_{S(t)}(\vb{x} \times \vb{u}) (\vb{u} - \vb{u}_s)\cdot\vu{n} \dd{S} + \oint_{S_b(t)}(\vb{x} \times \vb{u}) (\vb{u} - \vb{u}_s)\cdot\vu{n} \dd{S},
\end{multline}  
Here $\vb{M}$ is the total moment acting on the immersed body, $\vb{u}_s$ is the velocity of a moving surface, $\vb{I}$ is the identity tensor, and $\vb{T} = \nu(\nabla \vb{u} + \nabla \vb{u}^T)$ is the viscous stress tensor. To proceed, we use an an identity taken from \cite{Wu1981},
\begin{equation}\label{eq:jcwu}
    \int_{V(t)} \vb{x} \times \vb{u} \dd{V} = -\frac{1}{2}\int_{V(t)} \abs{\vb{x}}^2\bm{\omega} \dd{V} + \frac{1}{2} \oint_{S(t)} \abs{\vb{x}}^2 \vu{n} \times \vb{u} \dd{S} - \frac{1}{2} \oint_{S_b(t)} \abs{\vb{x}}^2 \vu{n} \times \vb{u} \dd{S}.
\end{equation} 
The left hand side represents the total angular momentum of the flow, while the right hand side represents the angular impulse along with boundary terms. Using (\ref{eq:jcwu}) to replace the first volume integral in (\ref{eq:angmom}) gives
\begin{multline}\label{impulsestarter}
 \vb{M} = \dv{}{t} \int_{V(t)} \frac{\abs{\vb{x}}^2}{2} \bm{\omega} \dd{V} - \dv{}{t} \oint_{S(t)} \frac{\abs{\vb{x}}^2}{2} \vu{n} \times  \vb{u} \dd{S} + \dv{}{t} \oint_{S_b(t)} \frac{\abs{\vb{x}}^2}{2} \vu{n} \times  \vb{u} \dd{S} \\
+ \oint_{S(t)} \vb{x} \times [(-p\vb{I} + \vb{T})\cdot \vu{n}] \dd{S}  - \oint_{S(t)}(\vb{x} \times \vb{u}) (\vb{u} - \vb{u}_s)\cdot\vu{n} \dd{S} + \oint_{S_b(t)}(\vb{x} \times \vb{u}) (\vb{u} - \vb{u}_s)\cdot\vu{n} \dd{S}.
\end{multline}
Taking a cue from Noca's derivation, we adopt the following program:
\begin{enumerate}
    \item Bring the time derivatives inside the integrals over the moving outer surface.
    \item Replace all time derivatives of velocity using the Navier Stokes equations.
    \item Transform any resulting terms containing $\nabla p$ into terms containing $p$ only. 
\end{enumerate}
These new pressure terms should exactly cancel the existing pressure term, leaving a pressure-free control volume formulation. To accomplish the first step, we need the tensor identity
\begin{equation} \label{tensortransport} 
\dv{}{t} \oint_{S(t)} \vb{A}\vu{n} \dd{S} = \oint_{S(t)} \pdv{\vb{A}}{t}\vu{n} \dd{S} + \oint_{S(t)} \nabla \cdot \vb{A} (\vb{u}_s\cdot \vu{n}) \dd{S}, 
\end{equation}
where $\vb{A}(\vb{x},t)$ is rank-two tensor field and $(\nabla \cdot \vb{A})_i = \sum_j \partial_jA_{ij}$. 
Let $[\vb{a}]$ be the cross-product matrix for a vector field $\vb{a}$, so that $[\vb{a}] \vb{x} = \vb{a} \times \vb{x}$. Applying (\ref{tensortransport}) to (\ref{impulsestarter}) gives
\begin{align}
    - \dv{}{t} \oint_{S(t)} \frac{\abs{\vb{x}}^2}{2} \vu{n} \times  \vb{u} \dd{S} &= \dv{}{t} \oint_{S(t)} \qty[\frac{\abs{\vb{x}}^2}{2} \vb{u}]\vu{n} \dd{S} \nonumber \\
    &= \oint_{S(t)} \pdv{}{t}\qty[\frac{\abs{\vb{x}}^2}{2} \vb{u}] \vu{n} \dd{S} + \oint_{S(t)} \nabla \cdot \qty[\frac{\abs{\vb{x}}^2}{2} \vb{u}] (\vb{u}_s\cdot \vu{n}) \dd{S}
\end{align}
Noting that $\nabla \cdot [\vb{a}] = -\nabla \times \vb{a}$,
\begin{align}
    &= \oint_{S(t)} \frac{\abs{\vb{x}}^2}{2} \pdv{\vb{u}}{t} \times \vu{n} \dd{S} - \oint_{S(t)} \nabla \times \qty(\frac{\abs{\vb{x}}^2}{2} \vb{u}) (\vb{u}_s\cdot \vu{n}) \dd{S} \nonumber \\
    &= -\oint_{S(t)} \frac{\abs{\vb{x}}^2}{2} \vu{n} \times \pdv{\vb{u}}{t} \dd{S} - \oint_{S(t)} \qty(\vb{x} \times \vb{u} + \frac{1}{2}\abs{\vb{x}}^2 \bm{\omega})(\vb{u}_s\cdot \vu{n}) \dd{S}. \label{interm1}
\end{align}
In the last step the vector identity $\vb{x}\times\vb{a} = -\frac{1}{2}\abs{\vb{x}}^2\nabla\times\vb{a} + \frac{1}{2} \nabla \times (\abs{\vb{x}}^2\vb{a})$  has been used. To eliminate the time derivative of velocity, we use the Navier Stokes equations in rotational form, 
$$\pdv{\vb{u}}{t} = -\nabla\qty(p + \frac{1}{2}\abs{\vb{u}}^2) + \vb{u}\times\bm{\omega} + \nabla \cdot \vb{T}. $$
Substituting this into the first term of (\ref{interm1}),
\begin{align}
  -\oint_{S(t)} \frac{\abs{\vb{x}}^2}{2} \vu{n} \times \pdv{\vb{u}}{t} \dd{S} &= -\oint_{S(t)} \frac{\abs{\vb{x}}^2}{2} \vu{n} \times \qty(-\nabla\qty(p + \frac{1}{2}\abs{\vb{u}}^2) + \vb{u}\times\bm{\omega} + \nabla \cdot \vb{T}) \dd{S} \nonumber \\
  &= \oint_{S(t)} \frac{\abs{\vb{x}}^2}{2} \vu{n} \times \nabla\qty(p + \frac{1}{2}\abs{\vb{u}}^2) \dd{S} \nonumber \\ & \quad \quad - \oint_{S(t)} \frac{\abs{\vb{x}}^2}{2} \vu{n}\times(\vb{u}\times\bm{\omega}) \dd{S} - \oint_{S(t)} \frac{\abs{\vb{x}}^2}{2} \vu{n}\times (\nabla \cdot \vb{T}) \dd{S}. \label{interm2}
\end{align}
The last two integrals on the right hand side do not involve the pressure, and will not be manipulated further. To transform the pressure gradient term in \eqref{interm2}, we use the integral identity 
\begin{equation}
    \oint_{S(t)} \frac{\abs{\vb{x}}^2}{2}\vu{n}\times\nabla\phi\dd{S} = \oint_{S(t)} \vb{x}\times \phi\vu{n}\dd{S}.
\end{equation} 
Applying this to the first integral in \eqref{interm2} gives.
\begin{align}
    \oint_{S(t)} \frac{\abs{\vb{x}}^2}{2} \vu{n} \times \nabla\qty(p + \frac{1}{2}\abs{\vb{u}}^2) \dd{S} &= \oint_{S(t)} \vb{x} \times \qty(p + \frac{1}{2}\abs{\vb{u}}^2) \vu{n} \dd{S} \nonumber \\
    &= \oint_{S(t)} \vb{x} \times p \vu{n} \dd{S} + \oint_{S(t)} \vb{x} \times \frac{1}{2}\abs{\vb{u}}^2 \vu{n} \dd{S} \label{interm3}.
\end{align} 
To collect these results, we substitute (\ref{interm3}) into (\ref{interm2}), then substitute (\ref{interm2}) into (\ref{interm1}), and finally substitute (\ref{interm1}) into (\ref{impulsestarter}). Canceling the pressure terms and collecting the surface terms brings us to an angular-impulse based control volume formula for moments,
\begin{equation}\label{impderivation}
 \vb{M} = \dv{}{t} \int_{V(t)} \frac{\abs{\vb{x}}^2}{2} \bm{\omega} \dd{V}  + \dv{}{t} \oint_{S_b(t)} \frac{\abs{\vb{x}}^2}{2} \vu{n} \times  \vb{u} \dd{S} + \oint_{S_b(t)}(\vb{x} \times \vb{u}) (\vb{u} - \vb{u}_s)\cdot\vu{n} \dd{S} + \oint_{S(t)} \bm{\lambda}(\vu{n}) \dd{S},
\end{equation}
where the quantity $\lambda(\vu{n})$ collects miscellaneous surface terms:
\begin{equation}
    \begin{aligned}
    \bm{\lambda}(\vu{n}) =  \vb{x} \times \frac{1}{2}\abs{\vb{u}}^2 \vu{n} &- \frac{\abs{\vb{x}}^2}{2} \vu{n}\times(\vb{u}\times\bm{\bm{\omega}}) - \frac{\abs{\vb{x}}^2}{2} \vu{n}\times (\nabla \cdot \vb{T}) - \qty(\frac{1}{2}\abs{\vb{x}}^2 \bm{\bm{\omega}})(\vb{u}_s\cdot \vu{n}) \\ &+ \vb{x}\times \qty(\vb{T}\cdot \vu{n}) - (\vb{x} \times \vb{u}) (\vb{u} \cdot\vu{n}).
    \end{aligned}
\end{equation}
We are now free to use (\ref{eq:jcwu}) to transform the above back into a formulation based on angular momentum. Doing so removes the integral over the immersed boundary, in exchange for an extra integration around the edge of the domain:
\begin{equation}\label{momderivation}
 \vb{M} = -\dv{}{t} \int_{V(t)} \vb{x} \times \vb{u} \dd{V}  + \dv{}{t} \oint_{S(t)} \frac{\abs{\vb{x}}^2}{2} \vu{n} \times  \vb{u} \dd{S} + \oint_{S_b(t)}(\vb{x} \times \vb{u}) (\vb{u} - \vb{u}_s)\cdot\vu{n} \dd{S} + \oint_{S(t)} \bm{\lambda}(\vu{n}) \dd{S}.
\end{equation}
This last equation, when specialized to a 2D stationary body, is the control volume moment formulation presented in \eqref{Mm1} and \eqref{Mm2}.

\end{document}